\documentclass[lettersize,journal]{IEEEtran}
\usepackage{amsmath,amsfonts}
\usepackage{algorithmic}
\usepackage{algorithm}
\usepackage{array}
\usepackage{booktabs}
\usepackage[caption=false,font=normalsize,labelfont=sf,textfont=sf]{subfig}
\usepackage{textcomp}
\usepackage{stfloats}
\usepackage{url}
\usepackage{verbatim}
\usepackage{graphicx}
\usepackage{cite}
\usepackage{hyperref}
\usepackage{wrapfig}

\usepackage{float}
\usepackage{tikz}
\usetikzlibrary{shapes,arrows, snakes}

\tikzstyle{specialblock} = [draw, ultra thick, fill=blue!20, rectangle, 
    minimum height=3em, minimum width=4em]
\tikzstyle{block} = [draw, fill=lightgray, rectangle, 
    minimum height=3em, minimum width=4em]
\tikzstyle{sum} = [draw, fill=white, circle, node distance=1cm]
\tikzstyle{prod}   = [circle, minimum width=8pt, draw, inner sep=0pt, path picture={\draw (path picture bounding box.south east) -- (path picture bounding box.north west) (path picture bounding box.south west) -- (path picture bounding box.north east);}]
\tikzstyle{sumt}   = [circle, minimum width=8pt, draw, inner sep=0pt, path picture={\draw (path picture bounding box.east) -- (path picture bounding box.west) (path picture bounding box.south) -- (path picture bounding box.north);}]
\tikzstyle{input} = [coordinate]
\tikzstyle{output} = [coordinate]
\tikzstyle{pinstyle} = [pin edge={to-,thin,black}]
\tikzset{
tmp/.style  = {coordinate}, 
dot/.style = {circle, minimum size=#1,
              inner sep=0pt, outer sep=0pt},
dot/.default = 6pt % size of the circle diameter 
}
\usepackage{amssymb}

\usepackage{mathtools}

\usepackage{capt-of}
\usepackage{url}
\usepackage{xparse,xfp}

\begin{document}
%\title{On Using Transformers for Speech Separation}
\title{Exploring Self-Attention Mechanisms \\for Speech Separation}


\author{Cem Subakan$^{1,2,3}$, Mirco Ravanelli$^{2,3}$, Samuele Cornell$^4$, Fran\c{c}ois Grondin$^5$, Mirko Bronzi$^3$ \\% <-this % stops a space
%\thanks{hello}% <-this % stops a space
%\thanks{all}% <-this % stops a space
$^1$Universit{\'e} Laval, $^2$Concordia University, $^3$Mila-Quebec AI Institute, $^4$Università Politecnica delle Marche, \\ $^5$Universit{\'e} de Sherbrooke. 
}

\markboth{Journal of \LaTeX\ Class Files,~Vol.~14, No.~8, August~2021}%
{Shell \MakeLowercase{\textit{et al.}}: A Sample Article Using IEEEtran.cls for IEEE Journals}

%\markboth{Journal of \LaTeX\ Class Files,~Vol.~14, No.~8, August~2015}%%{Shell \MakeLowercase{\textit{et al.}}: Bare Demo of IEEEtran.cls for IEEE Journals}



\usetikzlibrary{arrows.meta}

\maketitle

\begin{abstract}
Transformers have enabled impressive improvements in deep learning. They often outperform recurrent and convolutional models in many tasks while taking advantage of parallel processing. Recently, we proposed the SepFormer, which obtains state-of-the-art performance in speech separation with the WSJ0-2/3 Mix datasets. This paper studies in-depth Transformers for speech separation. In particular, we extend our previous findings on the SepFormer by providing results on more challenging noisy and noisy-reverberant datasets, such as LibriMix, WHAM!, and WHAMR!. Moreover, we extend our model to perform speech enhancement and provide experimental evidence on denoising and dereverberation tasks. Finally, we investigate, for the first time in speech separation, the use of efficient self-attention mechanisms such as Linformers, Lonformers, and ReFormers. We found that they reduce memory requirements significantly. For example, we show that the Reformer-based attention outperforms the popular Conv-TasNet model on the WSJ0-2Mix dataset while being faster at inference and comparable in terms of memory consumption.
\end{abstract}

\begin{IEEEkeywords}
speech separation, source separation, transformer, attention, deep learning.
\end{IEEEkeywords}

% \fi
\IEEEpeerreviewmaketitle

\newcommand{\mixture}{x}
\newcommand{\ldim}{F}
\newcommand{\len}{T}
\newcommand{\llen}{T'}
\newcommand{\nspk}{Ns}
\newcommand{\nsepf}{M}
\newcommand{\chnksize}{C}
\newcommand{\hopsize}{H}
\newcommand{\numchnks}{Nc}
\newcommand{\numintra}{N_\text{intra}}
\newcommand{\numinter}{N_\text{inter}}

\newcommand{\cem}[1]{#1}
\newcommand{\cemAQ}[1]{#1}

\section{Introduction}
\IEEEPARstart{T}{ransformers} are playing a pivotal role in modern deep learning \cite{vaswani2017}. They contributed to a paradigm shift in sequence learning and made it possible to achieve unprecedented performance in many Natural Language Processing (NLP) tasks, such as language modeling \cite{devlin-etal-2019-bert}, machine translation \cite{liu-etal-2020-multilingual-denoising}, and various applications in computer vision \cite{dosovitskiy2021image, gong2021vision}.
Transformers enable more accurate modeling of longer-term dependencies, which makes them suitable for speech and audio processing as well.
They have indeed been recently adopted for speech recognition, speaker verification, speech enhancement, and other tasks \cite{dong2018speechtransformer, gulati2020conformer}. 

\cem{Powerful sequence modeling is central to source separation as well, where long-term modeling turned out to impact performance significantly \cite{luo2020dualpath, dptn}.
However, speech separation typically involves long sequences in the order of tens of thousands of frames, and using Transformers poses a challenge because of the quadratic complexity of the standard self-attention mechanism \cite{vaswani2017}. 
For a sequence of length $N$, it needs to compare $N^2$ elements, leading to a computational bottleneck that emerges more clearly when processing long signals}. Mitigating the memory bottleneck of Transformers has been the object of intense research in the last years \cite{child2019generatingsparse, beltagy2020longformer, wang2020linformer, kitaev2020reformer}.
A popular way to address the problem consists of attending to a subset of elements only. For instance, Sparse Transformer \cite{child2019generatingsparse} employs local and dilated sliding windows over a fixed number of elements to learn short and long-term dependencies. LongFormer \cite{beltagy2020longformer} augments the Sparse Transformer by adding global attention heads. Linformers \cite{wang2020linformer}, approximate the full sparse attention with low-rank matrix multiplication, while Reformers \cite{kitaev2020reformer} cluster the elements to attend through an efficient locality-sensitive hashing function. 

To address this issue in speech separation, we recently proposed the SepFormer \cite{subakan2020attention}, a Transformer specifically designed for speech separation that inherits the dual-path processing pipeline proposed originally in \cite{luo2020dualpath} for recurrent neural networks (RNN).
%especially for speech separation, where long-term modeling has been shown to impact performance significantly \cite{luo2020dualpath, dptn}. 
%Motivated by this reason, we proposed the Separation Transformer-\emph{SepFormer} \cite{subakan2020attention}, a transformer-based model that obtains state-of-the-art results on speech separation. 
% to address reviewer comment
%\cem{Powerful sequence modeling appears to be central to source separation as shown by \cite{luo2018convtasnet, luo2020dualpath}. However, in such application, which typically involves long sequences in the order of tens of thousands frames,  using Transformers directly poses a challenge because of the quadratic complexity of the regular self-attention \cite{vaswani2017}. }
%SepFormer makes use of the the dual-path processing pipeline proposed  originally in \cite{luo2020dualpath} for recurrent neural networks (RNN). 
Dual-path processing splits long sequences into chunks, thus naturally alleviating the quadratic complexity issue. Moreover, it combines short-term and long-term modeling by adding a cascade of two distinct stages, making it particularly appealing for audio processing. Differently from its predecessors, SepFormer employs a Masking Network composed of Transformer encoder layers only. This property significantly improves the parallelization and inference speed of the SepFormer compared to previous RNN-based methods.

%In this paper, building on top of this first work where we proposed SepFormer, we study to what extent transformer architectures are a viable choice for speech separation and speech enhancement applications. 
This paper extends our previous studies on Transformer-based architectures for speech separation. In our previous work \cite{subakan2020attention}, for instance, we focused on the standard WSJ0-2/3Mix benchmark datasets only.
Here, we propose additional experiments and insights on more realistic and challenging datasets such as  Libri2/3-Mix \cite{cosentino2020librimix}, which includes long mixtures, WHAM! \cite{wichern2019wham} and WHAMR! \cite{maciejewski2020whamr}, which feature noisy and noisy and reverberant conditions, respectively. 
Moreover,  we adapted the SepFormer to perform speech enhancement and provide experimental evidence on VoiceBank-DEMAND, WHAM!, and WHAMR! datasets.
% Moreover, to assess generalization of SepFormer on real-life mixtures, we use the recently released REAL-M dataset \cite{subakan2021realm}, which is composed of real-world speech mixtures. 
%In addition to these standard benchmark datasets we also provide results  %
Another contribution of this paper is investigating different types of self-attention mechanisms for speech separation.
We explore, for the first time in speech separation, three efficient self-attention mechanisms \cite{luo2020dualpath}, namely Longformer \cite{beltagy2020longformer}, Linformer \cite{wang2020linformer}, and Reformer \cite{kitaev2020reformer} attention, all found to yield comparable or even better results than standard self-attention in natural language processing (NLP) applications.

We found that Reformer and Longformer attention mechanisms exhibit a favorable trade-off between performance and memory requirements. For instance, the Reformer-based model turned out to be even more efficient in terms of memory usage and inference speed than the popular Conv-TasNet model while yielding significantly better performance than the latter (16.7\,dB versus 15.3\,dB SI-SNR improvement). \cem{Despite the impressive improvements in the memory footprint and inference time, we found that the best performance, in speech separation, is still achieved, by far, with the regular self-attention mechanism used in the original SepFormer model, confirming the importance of full self-attention mechanisms.}
%We also conclude after this ablation study with different types of self attention, (including the Conformer block \cite{gulati2020conformer}) that the best performance is obtained with regular self-attention mechanism that is used in the original SepFormer model}.

\cem{The training recipes for the main experiments are available in the Speechbrain \cite{speechbrain} toolkit. Moreover, the pretrained models for the WSJ0-2/3Mix datasets, Libri2/3Mix datasets, WHAM!/WHAMR! datasets are available on the SpeechBrain Huggingface page\footnote{\url{https://huggingface.co/speechbrain}}.}

\cemAQ{The contributions in this paper are summarized as follows, 
\begin{itemize}
    \item We extend the experimental results obtained with SepFormer to include Libri2/3Mix, WHAM!, and WHAMR! datasets. We also include speech enhancement results obtained with SepFormer on Voicebank-DEMAND, WHAM!, and WHAMR! datasets. 
    \item We investigate using efficient self-attention mechanisms such as Longformer, Linformer, and Reformer attention mechanisms within SepFormer. 
\end{itemize}
}

The paper is organized as follows. In Section \ref{sec:transf}, we introduce the building blocks of Transformers and also describe the three efficient attention mechanisms (Longformer, Linformer, and Reformer) explored in this work. In Section \ref{sec:sepformer}, we describe the SepFormer model in detail with a focus on the dual-path processing pipeline employed by the model. Finally, in Section \ref{sec:experiments}, we present the results on speech separation, speech enhancement, and experimental analysis of different types of self-attention.

%In this paper, we extend the results obtained in our original paper \cite{subakan2020attention}, to include results on reverberant and noisy settings via the WHAM! \cite{wichern2019wham}, and WHAMR! \cite{maciejewski2020whamr} datasets. We also include results from Libri2/3-Mix Datasets that include audio mixtures that include longer and more diverse utterances. We observe that SepFormer is able to obtain state-of-the art results on the majority of the datasets, while being computationally rather advantegous compared to some of the popular models in the litterature.  

%In addition, we also provide ablations on different types of self-attention mechanisms that have recently been proposed \cite{tay2020efficient, beltagy2020longformer, wang2020linformer, kitaev2020reformer}. We analyze the performance as well as the computational advantages of each variant. 

%Recently there has been a variety of papers which proposed different self-attention mechanisms to reduce the quadratic complexity of the self-attention block . In this paper, we investigate adopting several of these newly proposed architectures for source separation to compare and contrast their advantages and shortcomings compared to SepFormer. 

%We also extend the results obtained to include more extensive experimental evaluation as well as an investigation over different transformer architectures which apply different types of self-attention. 
\section{Related Works}

{\subsection{Deep learning for source separation}}

\cem{Thanks to the advances in deep learning, tremendous progress has been made in the source separation domain. Notable early works include Deep Clustering \cite{hershey2015deep}, where a recurrent neural network is trained on an affinity matrix in order to estimate embeddings for each source from the magnitude spectra of the mixture. TasNet and Conv-TasNet \cite{luo2017tasnet, luo2018convtasnet} achieved impressive performance by introducing time-domain processing, combined with utterance level and permutation-invariant training \cite{kolbaek2017multitalker}.}

\cem{
More recent works include Sudo rm -rf \cite{tzinis2020sudo}, which uses a U-Net type architecture to reduce computational complexity, and 
Dual-Path Recurrent Neural Network (DPRNN) \cite{luo2020dualpath}. The latter first introduced the extremely effective dual-path processing framework adopted in this work.
Other works focused more on training strategies, such as \cite{tzinis2019twostep}, which proposed to split the learning of the encoder-decoder pair and the masking network into two steps in time domain separation. 
\cite{nachmani2020voice} reports impressive performance with a modified DPRNN model using a technique to determine the number of speakers. In \cite{zeghidour2020wavesplit} Wavesplit was proposed and it further improved over \cite{nachmani2020voice} by leveraging speaker-id information.}

\subsection{Transformers for source separation}
At the time of the writing, only a few papers exist on Transformers for source separation. 
These works include DPTNet \cite{dptn}, which adopts an architecture similar to DPRNN but adds a multi-head-attention layer plus layer normalization in each block before the RNN part. SepFormer, on the contrary, uses multiple Transformer encoder blocks without any RNN in each dual-path processing block. 
Our results in Section \ref{sec:experiments} indicate that SepFormer outperforms DPTNet, and is faster and more parallelizable at inference time due to the absence of recurrent operations. 

In \cite{chen2020dont}, the authors propose Transformers to deal with a multi-microphone meeting-like scenario,  where the amount of overlapping speech is low and a powerful separation might not be needed. The system's main feature is adapting the number of transformer layers according to the complexity of the input mixture.
In \cite{zhao2020MSGtransformer}, a multi-scale transformer  is proposed. As shown in Section \ref{sec:experiments}, the SepFormer outperforms this method. 
\cem{Other Transformer-based source separation works include \cite{zhangzining2021}, which uses a similar architecture as SepFormer but uses RNNs before the self-attention blocks (like DPTNet).}
\cem{There also exists methods that leverage Transformers for continuous speech separation, which aim to do source separation on long, meeting-like audio recordings. Prominent examples include \cite{chensanyuan2021, chen2020continuous, chen2020dont}. }
\cemAQ{Moreover, there have been recent concurrent works exploring efficient attention mechanisms for speech separation \cite{performerinse, luo22_interspeech}. We note that, the architectures of the separation networks that we explore for this purpose are different from these aforementioned works.}

\tikzstyle{timeblocktrenc} = [draw, thick, fill=blue!10, rectangle, 
    minimum height=1.9em, minimum width=7.2cm, ] 

\section{Transformers}
\label{sec:transf}

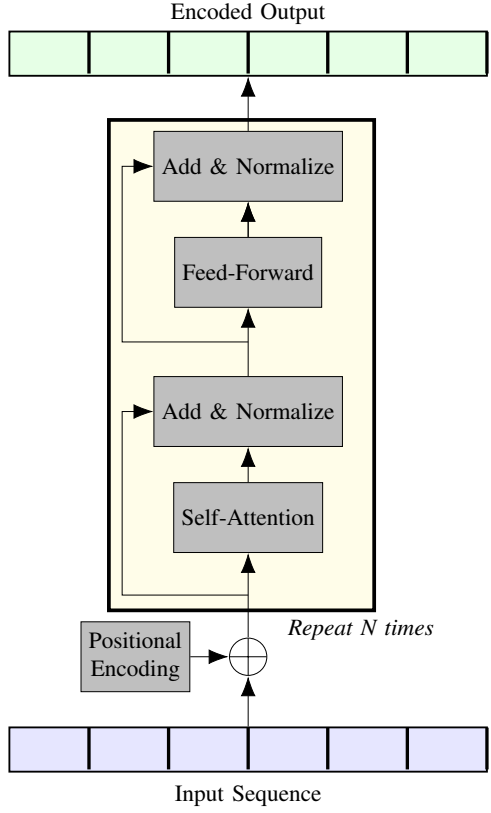
\begin{figure}[t!]
%\centering
%\center{\includegraphics[width=0.3\linewidth]{Transformer.pdf}}

\label{fig:transformer}

    \begin{center}
    \resizebox{6.6cm}{!}{
    \begin{tikzpicture}[auto, node distance=1.0cm,>=latex']
    \hspace*{-0.04\linewidth}
        
        \node [draw, ultra thick, fill=yellow!10, rectangle, xshift=-6.5cm, yshift=5.8cm, minimum width= 4.0cm, minimum height= 7.4cm] (plate) {};
        \node [draw=none, below of=plate, yshift=-3.0cm, xshift=1.8cm] (platetext) {\textit{Repeat N times}};
        %\node [draw=none, below of=plate,yshift=1cm, xshift=3.0cm] (platetext) {\textit{Repeat N times}};
        
        \node [timeblocktrenc, fill, xshift=-6.4cm] (input) {};  
        \node [below of=input, yshift=.3cm] (inptext) {Input Sequence};
        
        \foreach \x in {0,...,5}{
            \draw [line width=0.5mm](-\x*1.2 - 2.8, -0.3) -- (-\x*1.2 - 2.8, 0.33);
        }
        
        \node[sumt, above of=input, scale=2, yshift=.2cm] (sum) {};
        \node[block, left of=sum, align=center, xshift=-.7cm] (posenc) {Positional\\Encoding};
        \node[block, above of=sum, yshift=1.1cm] (selfat) {Self-Attention}; 
        \node[block, above of=selfat, yshift=.6cm] (AddNorm) {Add \& Normalize}; 
        \node[block, above of=AddNorm, yshift=1.1cm] (FeedForward) {Feed-Forward}; 
        \node[block, above of=FeedForward, yshift=.6cm] (AddNorm2) {Add \& Normalize}; 
        
        \node [timeblocktrenc, fill, fill=green!10, xshift=0cm, yshift=.7cm, above of=AddNorm2] (output) {};  
        \node [above of=output, yshift=-.4cm] (inptext) {Encoded Output};
        
        \foreach \x in {0,...,5}{
            \draw [line width=0.5mm](-\x*1.2 - 2.8, -0.3 + 10.5) -- (-\x*1.2 - 2.8, 0.33 + 10.5);
        }
        
        \draw [-{Latex[length=3mm]}] (input) -- (sum);
        \draw [-{Latex[length=3mm]}] (posenc) -- (sum);
        \draw [-{Latex[length=3mm]}] (sum) -- node (mid1) {} (selfat);
        \draw [-{Latex[length=3mm]}] (selfat) -- (AddNorm);
        \draw [-{Latex[length=3mm]}] (AddNorm) -- node (mid2) {} (FeedForward);
        \draw [-{Latex[length=3mm]}] (FeedForward) -- (AddNorm2);
        \draw [-{Latex[length=3mm]}] (FeedForward) -- (AddNorm2);
        \draw [-{Latex[length=3mm]}] (AddNorm2) -- (output);
        
        \node[left of=selfat, xshift=-.9cm] (g1) {};
        \draw [-{Latex[length=3mm]}] (mid1.east) -| (g1.center) |- (AddNorm.west);
        
        \node[left of=FeedForward, xshift=-.9cm] (g2) {};
        \draw [-{Latex[length=3mm]}] (mid2.east) -| (g2.center) |- (AddNorm2.west);
        %\draw [->] (zinp) |- (tmp4) -| (sum3);
        
    \end{tikzpicture}
    }
    \end{center}

\caption{The encoder of a standard Transformer. Positional embeddings are added to the input sequence. Then, N encoding layers based on multi-head self-attention, normalization, feed-forward transformations, and residual connections process it to generate an output sequence.}
\end{figure}

In this paper, we utilize the encoder part of the Transformer architecture, which is depicted in Fig. 1. The encoder turns an input sequence $X=\{x_1,\dots,x_{\len}\} \in \mathbb R^{\ldim \times \len}$ \cem{(where $\ldim$ denotes the number of features, and $\len$ denotes the signal length)} into an output sequence $Y=\{y_0,\dots,y_{\len}\} \in \mathbb R^{\ldim \times \len}$ using a pipeline of computations that involve positional embedding, multi-head self-attention, normalization, feed-forward layers, and residual connections.

\subsection{Multi-head self-attention}
The multi-head self-attention mechanism allows the Transformer to model dependencies across all the elements of the sequence. The first step calculates the Query, Key, and Value matrices from the input sequence $X$ of length $\len$. This operation is performed by multiplying the input vector by weight matrices: $Q=W_QX$, $K=W_KX$, $V=W_VX$, where $W_Q, W_K, W_V \in \mathbb R^{d_\text{model} \times \ldim }$. The attention layer consists of the following operations:
\begin{equation}
 \text{Attention}(Q, K, V) = \text{SoftMax}\left (\frac{Q^\top K}{\sqrt{d_k}} \right ) V^\top, 
\end{equation}
where $d_k$ represents the latent dimensionality of the $Q$ and $K$ matrices. The attention weights are obtained from a scaled dot-product between queries and keys: closer queries and key vectors will have higher dot products. The softmax function ensures that the attention weights range between 0 and 1. The attention mechanism produces $\len$ weights for each input element  (i.e., $\len^2$ attention weights). All in all, the softmax part of the attention layer computes relative importance weights, and then we multiply this attention map with the input value sequence $V$. 
Multi-head attention consists of several parallel attention layers. More in detail, it is computed as follows: 
\begin{align}
    &\text{MultiHeadAttention}(Q, K, V) \\
    &=\text{Concat}(\text{Head}_1, \dots, \text{Head}_h ) W^O, \notag \\
    &\text{where Head}_i = \text{Attention}(Q_i, K_i, V_i), \notag
\end{align}

where $h$ is the number of parallel attention heads and $W^O \in \mathbb R^{d_\text{model} \times h \ldim }$ is the matrix that combines the parallel attention heads, and $d_\text{model}$ denotes the latent dimensionality of this combined model.

%The self-attention mechanism is replicated multiple times to model various dependencies across elements.  

\subsection{Feed-Forward Layers}
The feed-forward component of the Transformer architecture consists of a two-layer perceptron. The exact definition of it is the following:
\begin{align}
    \text{Feed-Forward}(x) = \text{ReLU}(x W_1 + b_1) W_2 + b_2, 
\end{align}
where $x$ is the input. In the context of sequences, this feed-forward transformation is applied to each time point separately. $W_1$, and $W_2$ are learnable matrices, and $b_1$, and $b_2$ are learnable bias vectors.

\subsection{Positional encoding}
As the self-attention layer and feed-forward layers do not embed any notion of order, the transformer uses a positional encoding scheme for injecting sequence ordering information. The positional encoding is defined as follows:
\begin{align}
    PE_{t, 2i} =& \text{Sin}(t / 10000^{2i/d_\text{model}}) \\
    PE_{t, 2i+1} =& \text{Cos}(t / 10000^{2i/d_\text{model}})
\end{align}

This encoding relies on sinusoids of different frequencies in each latent dimension to encode positional information \cite{vaswani2017}. 

%Many alternative approaches have been proposed in the literature such as \cite{}. Nevertheless, we found that absolute encoding leads to a better performance in speech separation.

\subsection{Reducing the memory bottleneck}
\label{sec:efficient-self-attention}
Many alternative architectures have been proposed in the literature. Popular variations for mitigating the quadratic memory issues are described in the following.

\subsubsection{Longformer} 

Longformer \cite{beltagy2020longformer} aims to reduce the quadratic complexity by replacing the full self-attention structure with a combination of local and global attention. 
%Specifically, Longformer utilizes sliding window attention to model local dependencies and global attention that enables a single token to attend to all of the elements. 
Specifically, Longformer relies on local attention that captures dependencies from nearby elements and global attention that globally captures dependencies from all the elements. To keep the computational requirements manageable, global attention is performed for a few special elements in the sequence.

%Specifically, Longformer relies on a local attention mechanism that captures dependencies only from nearby elements, and a global attention mechanism that globally captures dependencies from all the elements. To keep the computational requirements manageable, global attention is only performed for a few special elements in the sequence.
%The local attention is performed over a sliding window to bind the memory requirements. 
%These attention structures can be seen in Figure 2 of the longformer paper. 

\subsubsection{Linformer}
Linformer \cite{wang2020linformer} avoids the quadratic complexity by reducing the size of the time dimension of the matrices $K, V \in \mathbb R^{d_\text{model} \times \len}$. This is done by projecting the time dimension $\len$ to a smaller dimensionality $k$ by using projection matrices $P, F \in \mathbb R^{\len \times k}$. The multi-head attention equation, therefore, becomes the following: 
\begin{align}
    \text{LFSelfAttention}\left (Q, K, V \right ) = \text{SoftMax}\left (\frac{Q^\top (K P)}{\sqrt{d_k}} \right) (V F)^\top,
\end{align}
which effectively limits the complexity of the matrix product between the softmax output and the $V$ matrix. 

\subsubsection{Reformer}
Reformer \cite{kitaev2020reformer} uses locality-sensitive hashing (LSH) to reduce the complexity of self-attention. LSH is used to find the vector pairs $(q,k), q \in Q, k \in K$ that are closer. The intuition is that those pairs will have a bigger dot product and contribute the most to the attention matrix. Because of this, the authors limit the attention computation to the close pairs $(q,k)$, while ignoring the others (saving time and memory). In addition, the Reformer, inspired by \cite{DBLP:journals/corr/GomezRUG17}, implements reversible Transformer layers that avoid the linear memory complexity scaling with respect to the number of layers. \cem{An example usage of Reformer in speech is in Text-to-Speech \cite{reformertts}}.
%\subsection{Transformers for speech separation}

\section{Separation Transformer (SepFormer)}
\label{sec:sepformer}

% In this section we explore different types of approaches of applying self-attention to the masking based end-to-end source separation architecture. We first start with the definition of the masking-based source separation architecture. 

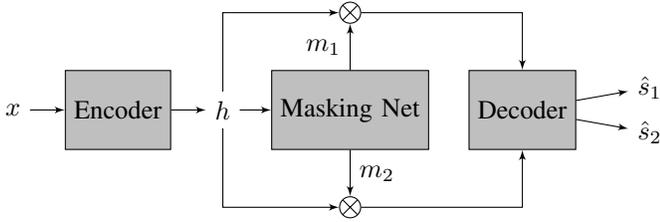
\begin{figure}[t!]
\centering
    \resizebox{9.2cm}{!}{
    \begin{tikzpicture}[auto, node distance=2.2cm,>=latex']
        \hspace*{-0.02\linewidth}
        \node [draw=none, fill=none] (input) {$\mixture$ };
        \node [block, right of=input, fill, xshift=-0.8cm ] (encoder) {Encoder};  
        \node [draw=none, fill=none, right of=encoder, xshift=-0.8cm] (latent) {$h$};
        \node [block, right of=latent, xshift=-0.5cm] (masknet) {Masking Net};
        \node [prod, above of=masknet,yshift=-0.9cm] (prod1) {}; 
        \node [prod, below of=masknet,yshift=0.9cm] (prod2) {}; 
        \node [block, right of=masknet, xshift=0.1cm] (decoder) {Decoder};
        \node [draw=none, fill=none, right of=decoder, yshift=0.3cm, xshift=-0.5cm] (out1) {$\hat{s}_1$};
        \node [draw=none, fill=none, right of=decoder, yshift=-0.3cm, xshift=-0.5cm] (out2) {$\hat{s}_2$};
        %\node [draw=none, fill=none, right of=masknet, yshift=0.1cm, xshift=-0.5cm] (mid1) {};
        %\node [draw=none, fill=none, right of=masknet, yshift=-0.1cm, xshift=-0.5cm] (mid2) {};
        
        \draw [->] (input) -- (encoder);
        \draw [->] (encoder) -- (latent);
        \draw [->] (latent) -- (masknet);
        \draw [->] (latent) |-  (prod1);
        \draw [->] (latent) |-  (prod2);
        \draw [->] (masknet) -- node {$m_1$} (prod1);
        \draw [->] (masknet) -- node {$m_2$} (prod2);
        \draw [->] (prod1) -| (decoder);
        \draw [->] (prod2) -| (decoder);
        %\draw [->] (mid1) -- (decoder);
        %\draw [->] (mid2) -- (decoder);
        \draw [->] (decoder) -- (out1);
        \draw [->] (decoder) -- (out2);
    \end{tikzpicture}
    }
\caption{The high-level description of the masking-based source separation pipeline: The encoder block learns a latent representation for the input signal. The masking network estimates optimal masks to separate the sources in the mixtures. The decoder finally reconstructs the estimated sources in the time domain using these masks. The self-attention-based modeling is applied inside the masking network.}
\label{fig:maskingpipeline}
\end{figure}

The SepFormer relies on the masking-based source separation framework popularized by \cite{luo2017tasnet, luo2018convtasnet} (Figure \ref{fig:maskingpipeline}). An input mixture $\mixture \in \mathbb R^{\len}$ feeds the architecture. An encoder learns a latent representation $h \in \mathbb R^{\ldim \times \llen}$. Afterwards, a masking network estimates the masks $m_1, m_2 \in \mathbb R^{\ldim \times \llen}$. The separated time domain signals for source 1 and source 2 are obtained by passing the $h*m_1$, and $h*m_2$ through the decoder, where $*$ denotes element-wise multiplication. In the following, we explain in detail each block separately.

\subsection{Encoder}
The encoder takes the time-domain mixture-signal $\mixture \in \mathbb R^T$ as input. It learns a time-frequency latent representation $h \in \mathbb R^{\ldim \times \llen}$ using a single convolutional layer:
\begin{align}
    h = \text{ReLU}(\text{Conv1d}(x)).  
\end{align}
As we will describe in Section \ref{sec:results-wsj0}, the stride factor of this convolution impacts significantly the performance, speed, and memory of the model.

\subsection{Decoder}
The decoder uses a simple transposed convolutional layer with the same stride and kernel size as the encoder. Its input is the element-wise multiplication between the mask of the  $k$-th source $m_k$ and the output of the encoder $h$. The transformation operated by the decoder is the following: 
\begin{align}
    \widehat {s}_k = \text{Conv1dTranspose}(m_k \odot h), 
\end{align}
%\vspace{-0.3cm}
where $\widehat{s}_k \in \mathbb R^T$ denotes the separated source $k$, and $\odot$ denotes element-wise multiplication.

\subsection{Masking Network}
\label{sec:maskingnetwork}
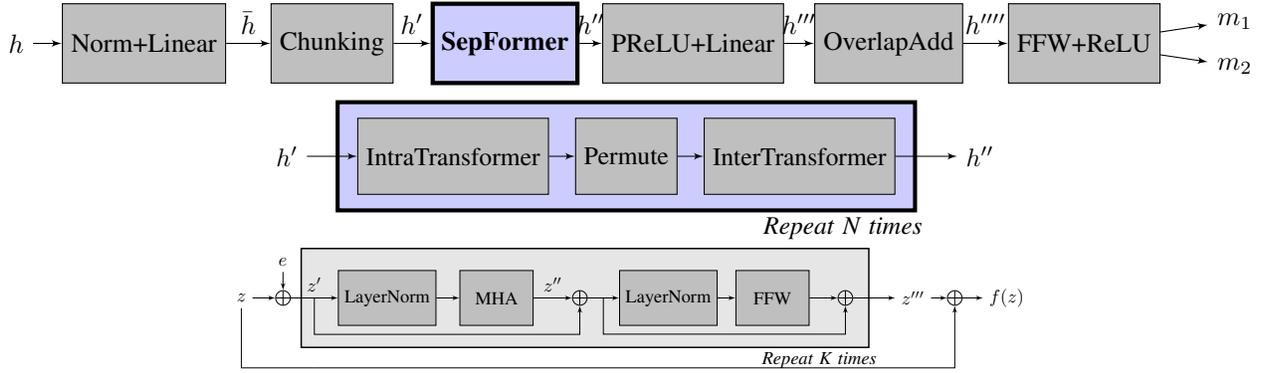
\begin{figure*}[t!]
\centering
    \newcommand{\sepfig}{0.5}
    \begin{tikzpicture}[auto, node distance=2.0cm,>=latex']
        \node [draw=none, fill=none] (h) {$h$};
        \node [block, right of=h, fill, xshift=-0.3cm] (1) {Norm+Linear};  
        \node [block, right of=1, fill, xshift=\sepfig cm] (2) {Chunking};  
        \node [specialblock, right of=2,  xshift=0.3 cm] (3) {\textbf{SepFormer}};  
        \node [block, right of=3, xshift=\sepfig cm] (4) {PReLU+Linear};
        \node [block, right of=4, xshift=0.6 cm] (5) {OverlapAdd};
        \node [block, right of=5, xshift=0.6 cm] (6) {FFW+ReLU};
        %\node [block, right of=2, fill, xshift=1cm] (4) {\textbf{SepFormer}};  
        \node [draw=none, fill=none, right of=6, yshift=0.3cm] (out1) {$m_1$};
        \node [draw=none, fill=none, right of=6, yshift=-0.3cm] (out2) {$m_2$};
        
        \draw [->] (h) -- (1);
        \draw [->] (1) -- node {$\bar h$} (2); 
        \draw [->] (2) -- node {$h'$} (3); 
        \draw [->] (3) -- node {$h''$} (4); 
        \draw [->] (4) -- node {$h'''$} (5); 
        \draw [->] (5) -- node {$h''''$} (6); 
        \draw [->] (6) -- (out1); 
        \draw [->] (6) -- (out2); 
    \end{tikzpicture}
    \vspace{0.2cm}
    
     \resizebox{10cm}{!}{    
    \begin{tikzpicture}[auto, node distance=2cm,>=latex']
        \node [draw, ultra thick, fill=blue!20, rectangle, right of=1, xshift=1.0cm, minimum width= 8.0cm, minimum height= 1.5cm] (plate) {};
        \node [draw=none, below of=plate,yshift=1cm, xshift=3.0cm] (platetext) {\textit{Repeat N times}};
        \node [draw=none, fill=none, xshift=0cm] (input) {$h'$};
        \node [block, right of=input, xshift=0.3cm] (1) {IntraTransformer};  
        \node [block, right of=1, xshift=0.4cm] (2) {Permute};
        \node [block, right of=2, xshift=0.4cm] (3) {InterTransformer};  
        \node [draw=none, right of=3, xshift=0.5cm] (out) {$h''$};
        
        \draw [->] (input) -- (1);
        \draw [->] (1) -- (2);
        \draw [->] (2) -- (3); 
        \draw [->] (3) -- (out);
    \end{tikzpicture}
    }
    
    \vspace{0.0cm}
   \resizebox{10.8cm}{!}{
    \begin{tikzpicture}[auto, node distance=2cm,>=latex']
        \node [draw, thick, fill=gray!20, rectangle, right of=1, xshift=2.3cm, minimum width= 10.9cm, minimum height= 1.9cm] (plate) {};
        \node [draw=none, below of=plate,yshift=0.8cm, xshift=4.5cm] (platetext) {\textit{Repeat K times}};
        \node [draw=none, ] (input) {$z$};
        \node [sumt, right of=input, xshift=-1.2cm] (sum) {};
        \node [draw=none, above of=sum, yshift=-1.3cm] (e) {$e$};
        \node [block, right of=sum] (1) {LayerNorm};
        \node [block, right of=2, xshift=-1.8cm] (2) {MHA};
        \node [sumt, right of=2, xshift=-0.4cm] (sum2) {};
        \node [block, right of=sum2, xshift=-0.3cm] (3) {LayerNorm};
        \node [block, right of=3] (4) {FFW};
        \node [sumt, right of=4, xshift=-0.6cm] (sum3) {};
        \node [draw=none, right of=sum3, xshift=-0.7cm] (out) {$z'''$};
        \node [sumt, right of=sum3, xshift=0.1cm] (sum4) {};
        \node [draw=none, right of=sum4, xshift=-1.0cm] (fout) {$f(z)$};
        \node [tmp, below of=1, yshift=1.3cm] (tmp1) {};
        \node [tmp, below of=sum2, yshift=1.3cm] (tmp2) {};
        
        \draw [->] (input) -- (sum);
        \draw [->] (e) -- (sum);
        \draw [->] (sum) -- node (zp) {$z'$} (1);
        \draw [->] (1) -- (2);
        \draw [->] (2) -- node (zpp) {$z''$}(sum2); 
        \draw [->] (sum2) -- node (zinp) {} (3);
        \draw [->] (3) -- (4);
        \draw [->] (4) -- (sum3);
        \draw [->] (sum3) -- (out);
        \draw [->] (zp) |-(tmp1) -| (sum2);
        %\draw [->] (tmp2) -| (sum3);
        \draw [->] (out) -- (sum4);
        \draw [->] (sum4) -- (fout);
        
        \node [tmp, below of=zinp, yshift=1.18cm] (tmp4) {};
        % for the missing residual connection
        \node [tmp, below of=zp, yshift=.4cm] (tmp3) {};
        \draw [-] (input) |- (tmp3);
        \draw[->] (tmp3) -| (sum4);
        \draw [->] (zinp) |- (tmp4) -| (sum3);
        
    \end{tikzpicture}
    }    
    \vspace{-0.3cm}
    \caption{\textbf{(Top)} The overall architecture proposed for the masking network. \textbf{(Middle)} The SepFormer Block.  \textbf{(Bottom)} The transformer architecture $f(.)$ that is used both in the IntraTransformer block and in the InterTransformer block.}
\label{fig:sepformer}
\end{figure*}

Figure \ref{fig:sepformer} (top) shows the detailed architecture of the masking network (Masking Net).
The masking network is fed by the encoded representations $h \in \mathbb R^{\ldim \times \llen}$ and estimates a mask $\{m_1,\dots,m_{\nspk}\}$ for each of the $\nspk$ speakers in the mixture. 

As in \cite{luo2018convtasnet}, the encoded input $h$ is normalized with layer normalization \cite{layernorm} and processed by a linear layer (with dimensionality $\ldim$). 
Following the dual-path framework introduced in \cite{luo2020dualpath}, we create overlapping chunks of size $\chnksize$ by chopping up $h$ on the time axis with an overlap factor of 50\%. We denote the output of the chunking operation with $h' \in \mathbb R^{\ldim \times \chnksize \times \numchnks}$, where $C$ is the length of each chunk, and $\numchnks$ is the resulting number of chunks. 

The representation $h'$ feeds the SepFormer separator block, which is the main component of the masking network. This block, which will be described in detail in Section \ref{sec:dppipeline-sepformer}, employs a pipeline composed of two Transformers able to learn short and long-term dependencies. \cemAQ{We pictorially describe the overall pipeline of the SepFormer separator block in Figure \ref{fig:dualpath}. We would like to note that, for graphical convenience, non-overlapping chunks were used in the figure. However, in the experiments in this paper, we used 50\% overlapping chunks.} 

The output of the SepFormer $h'' \in \mathbb R^{\ldim \times \chnksize \times \numchnks}$ is processed by PReLU activations followed by a linear layer.  
We denote the output of this module by $h''' \in \mathbb R^{(\ldim \times \nspk) \times \chnksize \times\numchnks}$, where $\nspk$ is the number of speakers. Afterward, we apply the overlap-add scheme described in \cite{luo2020dualpath} and obtain $h''''\in \mathbb R^{\ldim \times \nspk \times T'}$. We pass this representation through two feed-forward layers and a ReLU activation at the end to obtain a mask $m_k$ for each one of the speakers.

\subsection{Dual-Path Processing Pipeline and SepFormer}
\label{sec:dppipeline-sepformer}
The dual-path processing pipeline \cite{luo2020dualpath} (shown in Fig. \ref{fig:dualpath}) enables a combination of short and long-term modeling while making long sequence processing feasible with a self-attention block. In the context of the masking-based end-to-end source separation shown in Fig. \ref{fig:maskingpipeline}, the latent representation $h \in \mathbb R^{\ldim \times \llen}$ is divided into overlapping chunks.

A Transformer encoder is applied to each chunk separately. We call this block IntraTransformer (IntraT) and denote it with $f_\text{intra}(.)$. This block effectively operates as a block diagonal attention structure, modeling the interaction between all elements in each chunk separately so that the attention memory requirements are bounded. 
Another transformer encoder is applied to model the inter-chunk interactions, i.e. between each chunk element. We call this block InterTransformer (InterT) and denote with $f_\text{inter}(.)$. 
Because this block attends to all elements that occupy the same position in each chunk, it effectively operates as a strided-attention structure. Mathematically, the overall transformation can be summarized as follows: 
\begin{align}
    h'' = f_\text{inter}( f_\text{intra} ( h') ). 
\end{align}
In the standard SepFormer model defined in \cite{subakan2020attention}, $f_\text{inter}$ and $f_\text{intra}$ are chosen to be full self-attention blocks. In Section \ref{sec:full-self-attention}, we describe the full self-attention block employed in SepFormer in detail. 

%In Section \ref{sec:efficient-self-attention}, we describe recently proposed efficient transformer architectures that can be used as an alternative to the full self attention block.  

% \begin{figure*}
%     \centering
%     \includegraphics[width=\textwidth]{dualpath_wattention.pdf}
%     \caption{The dual-path processing pipeline that is employed in SepFormer.}
%     \label{fig:dualpath}
% \end{figure*}

    \tikzstyle{chunks} = [draw, thick, fill=blue!10, rectangle, 
    minimum height=4.2cm, minimum width=2cm] 
    
    \tikzstyle{timeblock} = [draw, thick, fill=blue!10, rectangle, 
    minimum height=1.9em, minimum width=14cm,  
    %node contents={
    %\tikz \node [] (a) {asd};  
     %\node [draw=none, below of=a, yshift=-1cm] (b) {dsa}; 
     %    } 
          ]
\begin{figure*}
    \centering
    \newcommand{\septime}{1}
    \newcommand{\yshift}{7.3}
    \resizebox{15cm}{!}{
    \begin{tikzpicture}[auto, node distance=1.0cm,>=latex']
    \hspace*{-0.04\linewidth}
        \node [xshift=-7.8cm, yshift=0cm] (hshow) {$\bar h$};
        \draw  [line width=0.6mm, -implies,double, double distance=.6mm] (hshow)-- node (taxis) [yshift=0.2cm] {chunk $\bar h$} (-6.5, 0);  
        
        \node [draw=none, fill=none] (h) {$h$};
        \node [timeblock, right of=h, fill, xshift=-0.3cm] (asd) {};  
        
        \foreach \x in {0,...,5}{
            \draw [line width=0.5mm](-\x*2 + 5.6, -0.5) -- (-\x*2 + 5.6, 0.5);
        }
        
        \foreach \x in {-3,...,3}{
            \node [xshift=2*\x cm](h\x) {$\bar h_{\fpeval{\x+3}}$};
        }
        
        \node [chunks, below of=h, yshift=-4.2cm, xshift=-5cm] (ch1) {};
        \foreach \y in {1,...,6}{
            \draw [line width=0.5mm](-4, 0.6*\y - \yshift) -- (-6, 0.6*\y-\yshift);
        }
        \foreach \y in {0,...,6}{
            \node [yshift=\fpeval{-0.59*\y-3.45} cm, xshift=-5.7cm](h\y) {$h^{'}_{\y}$};
        }
        
        \node [chunks, right of=ch1, xshift=5cm, fill=yellow!10] (ch2) {};
        \foreach \y in {1,...,6}{
            \draw [line width=0.5mm](0, 0.6*\y - \yshift) -- (2, 0.6*\y-\yshift);
            
        }
        \foreach \y in {1,...,7}{
            \draw [->, line width=0.4mm] (0.4, 0.6*\y - \yshift -0.3) -- (1.6, 0.6*\y-\yshift -0.3);
        }
        
        \node [chunks, right of=ch2, xshift=5cm, fill=red!10] (ch3) {};
        \foreach \y in {1,...,6}{
            \draw [line width=0.5mm](6, 0.6*\y - \yshift) -- (8, 0.6*\y-\yshift);
        }
        \foreach \x in {0,...,4}{
            \draw [->, line width=0.5mm, color=red](\x*0.4 + 6.2, -3.3) -- (\x*0.4 + 6.2, -7.2);
        }

        %\foreach \arrowtipkind[count=\i from 0] in {
        %Triangle}{\foreach \specs[count=\j from 0] in {round, open, fill=red, {round, fill=blue, length=2.5mm, slant=.5}}{\draw[-{\arrowtipkind[\specs]}, yshift=-1.5*\i cm -0.2*\j cm] (0,0) -- +(1,0)\ifnum\j=0 node[above,midway,font=\scriptsize\ttfamily]{\arrowtipkind}\fi;};};
        %\foreach \arrowtipkind[count=\i from 0] in {
        %Triangle}{\foreach \specs[count=\j from 0] in {round, open, fill=red, {round, fill=blue, length=2.5mm, slant=.5}}{\draw[-{\arrowtipkind[\specs]}, yshift=-1.5*\i cm -0.2*\j cm] (0,0) -- +(1,0)\ifnum\j=0 node[above,midway,font=\scriptsize\ttfamily]{\arrowtipkind}\fi;};};
        %\draw [line width=1mm, -implies,double, double distance=1mm] (-2,1)-- node (taxis) [yshift=0.2cm] {the time axis} (1,1); 
        \draw [line width=1mm, -implies,double, double distance=0mm] (-2,1)-- node (taxis) [yshift=0.2cm] {the time axis} (1,1); 
        
        \draw [line width=1mm, -implies,double, double distance=1mm] (-6,-0.8)-- node (taxis) [yshift=0.2cm, xshift=0.2cm] {concatenate chunks} (-6, -2.5); 
        
        %\draw [line width=.5mm, -implies,double, double distance=0.4mm, yshift=0.3cm, xshift=0.3cm] (-6.5,-5)-- node (taxis) [xshift=-1.8cm, text width=2cm] {the chunk axis} (-6.5, -6); 
        \draw [line width=.5mm, -implies,double, double distance=0.0mm, yshift=0.3cm, xshift=0.3cm] (-6.5,-5)-- node (taxis) [xshift=-1.8cm, text width=2cm] {the chunk axis} (-6.5, -6); 
        
        %\draw [line width=.5mm, -implies,double, double distance=0.4mm] (-5,-3)-- node (taxis) [xshift=-0.0cm, text width=2cm] {the time axis} (-4, -3); 
        \draw [line width=.5mm, -implies,double, double distance=0.0mm] (-5,-3)-- node (taxis) [xshift=-0.0cm, text width=2cm] {the time axis} (-4, -3); 
        
        \draw [line width=1mm, -implies,double, double distance=1mm, yshift=0.3cm] (-3.2, -5.5)-- node (taxis) [yshift=0.2cm] {IntraTransformer} (-1.2, -5.5); 
        
        \node [draw=none, fill=none, above of=taxis] (bdiag)  { \includegraphics[scale=0.1]{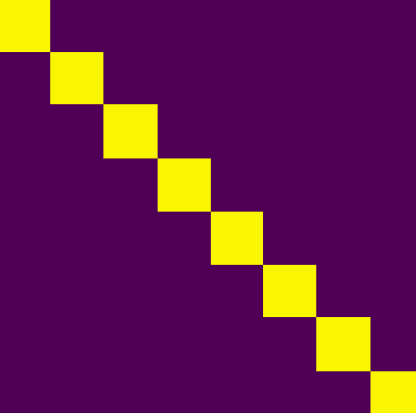}} ;

        \draw [line width=1mm, -implies,double, double distance=1mm, yshift=0.3cm] (2.9, -5.5)-- node (taxis) [yshift=0.2cm] {InterTransformer} (4.9, -5.5); 
        
        \node [draw=none, fill=none, above of=taxis] (dilated)  { \includegraphics[scale=0.1]{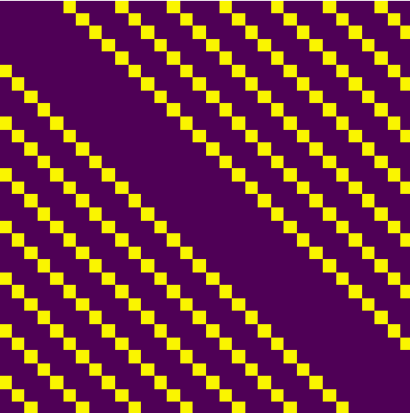}} ;
        
        %\draw[-{\arrowtipkind[round]}]
        
%%% Tips with particular options:
% Arc Barb[sep, arc=<angle>, length=<dim>, line width=<dim>, width=<dim>, reversed, round, slant=<num>, harpoon, left, right, <color>]
% Bracket[sep, reversed, round, slant=<num>, left, right, harpoon, reversed, <color>]
% Hooks[sep, arc=<angle>, length=<dim>, line width=<dim>, width=<dim>, reversed, round, slant=<num>, harpoon, left, right, <color>]
% Tee Barb[sep, inset=<dim>, inset'=<dim> <num>, line width=<dim>, reversed, round, slant=<num>, harpoon, left, right, <color>] thin thick
% Implies[<color>]
        
    \end{tikzpicture}
    }
    \caption{The dual-path processing employed in the SepFormer masking network. The input representation $\bar h$ is first of all chunked to get the chunks $\bar h_0$, $\bar h_1$, $\dots$, $\bar h_6$. Then the chunks are concatenated on another dimension (chunk dimension). Afterward, we apply the IntraTransformer along the time dimension of each chunk and the InterTransformer along the chunk dimension. \cemAQ{As mentioned in \ref{sec:maskingnetwork}, for graphical convenience, non-overlapping chunks are used in the figure. However, for the experiments in this paper, we used chunks with 50\% overlap.} The attention maps show how elements of the input feature vectors are interconnected sequentially inside the Intra and Inter blocks.    
    %The attention maps for the IntraTransformer and InterTransformer show the sequences elements which are inter-connected through the respective blocks.
    }
    \label{fig:dualpath}
\end{figure*}
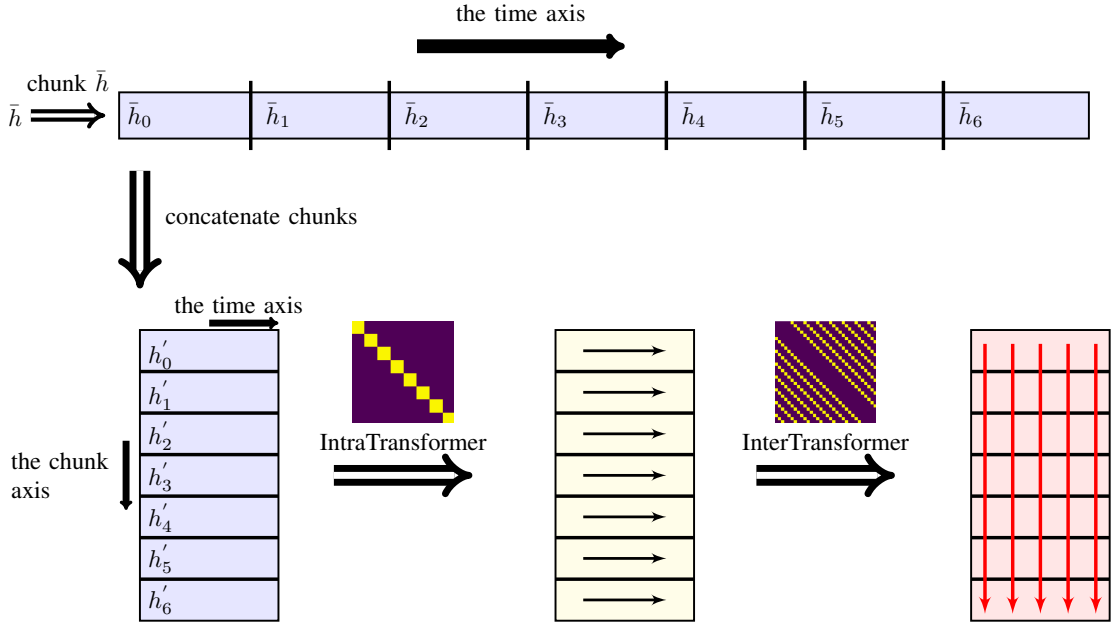

\subsection{Full Self-Attention Transformer Encoder}
\label{sec:full-self-attention}

The architecture for the full self-attention transformer encoder layers follows the original Transformer architecture \cite{vaswani2017}. Figure \ref{fig:sepformer} (Bottom) shows the architecture used for the IntraTransformer and InterTransformer blocks. 

We use the variable $z$ to denote the input to the Transformer. First of all, sinusoidal positional encoding $e$ is added to the input $z$, such that, 
\begin{align}
    z' = z + e.
\end{align}
We follow the positional encoding definition in \cite{vaswani2017}. 
We then apply multiple Transformer layers. 
Inside each Transformer layer $g(.)$, we first apply layer normalization, followed by multi-head attention (MHA):
\begin{align}
    z''= \text{MultiHeadAttention}(\text{LayerNorm}(z')).
\end{align}
As proposed in \cite{vaswani2017}, each attention head computes the scaled dot-product attention between all the sequence elements.
The Transformer finally employs a feed-forward network (FFW), which is applied to each position independently:
\begin{align}
    z''' = \text{Feed-Forward}(\text{LayerNorm}(z''+ z') ) + z'' + z'.
    \label{eq_ff}
\end{align}
The overall transformer block is therefore defined as follows:
\begin{align}
    f(z) =  g^K (z+e) + z, 
    \label{eq_intra}
\end{align}
where $g^K(.)$ denotes $K$ layers of transformer layer $g(.)$. We use $K=\numintra$ layers for the IntraT, and $K=\numinter$ layers for the InterT.
As shown in Figure \ref{fig:sepformer} (Bottom) and Eq. \eqref{eq_intra}, we add residual connections across the transformer layers and across the transformer architecture to improve gradient backpropagation. \cemAQ{We would like to note that, in the experiments presented in this paper, a non-causal attention mechanism was utilized.}

\section{Experimental setup}
\label{sec:experiments}

%First of all, we provide experiments to compare the performance different efficient architectures. For this, we use the popular bench-marking dataset WSJ0-2Mix. Afterwards, we provide results on various other datasets including WSJ0-3Mix, Libri2/3 Mix, WHAM!, and WHAMR! with SepFormer, extending upon our work in \cite{subakan2020attention}. 

\subsection{Datasets} 
In our experiments, we use the popular WSJ0-2mix and WSJ0-3mix datasets \cite{hershey2015deep}, where mixtures of two speakers and three speakers are created by randomly mixing utterances in the WSJ0 corpus. The relative levels for the sources are sampled uniformly between 0\,dB to 5\,dB. Respectively, 30, 10, and 5 hours of speech are used for training (20k utterances), validation (5k utterances), and testing (3k utterances). The training and test sets are created with different sets of speakers. We use the 8kHz version of the dataset, with the `min' version where the added waveforms are clipped to the shorter signal's length. These datasets have become the de-facto standard benchmark for source separation algorithms.

In addition to the WSJ0-2/3 Mix we also provide experimental evidence on WHAM! \cite{wichern2019wham}, and WHAMR! datasets \cite{maciejewski2020whamr} which are essentially derived from the WSJ0-2Mix dataset by adding environmental noise and environmental noise plus reverberation respectively. \cem{In WHAM! and WHAMR! datasets the environmental noises contain ambient noise from coffee shops, restaurants, and bars, which is collected by the authors of the original paper \cite{wichern2019wham}. In WHAMR! the reverberation time RT60 is randomly uniformly sampled from three categories--low, medium, and high, which respectively have the distributions $\mathcal U(0.1, 0.3)$, $\mathcal U(0.2, 0.4)$, $\mathcal U(0.4, 1.0)$. More details regarding the $RT_{60}$ distribution of the WHAMR! dataset is given in the original paper \cite{maciejewski2020whamr}.  }

The popular speech enhancement VoiceBank-DEMAND dataset \cite{ValentiniBotinhao2016InvestigatingRS} is also used to compare the SepFormer architecture with other state-of-the-art denoising models.
We also provide experimental results on the LibriMix dataset \cite{cosentino2020librimix}, which contains longer and more challenging mixtures than the WSJ0-Mix dataset. 

% We can add more details for LibriMix here. 

%
\subsection{Architecture Details}

The SepFormer encoder employs 256 convolutional filters with a kernel size of 16 samples and a stride factor of 8. The decoder uses the same kernel size and the stride factors of the encoder. 

In our best models, the masking network processes chunks of size $\chnksize=250$ with a 50\% overlap between them and employs 8 layers of Transformer encoder in both IntraTransformer and InterTransformer. The dual-path processing pipeline is repeated $N=2$ times. We used 8 parallel attention heads and 1024-dimensional positional feed-forward networks within each Transformer layer. The model has a total of 25.7 million parameters. 

\subsection{Training Details}
\newcommand{\dataaug}{DM}
%Training is performed with the WSJ0-2mix or  WSJ0-3mix data without using on-the-fly mixture generation to create unseen speaker combinations. Data contamination with noise or reverberation \cite{contamiation} is not used as well.
We use dynamic mixing (DM) data augmentation \cite{zeghidour2020wavesplit}, which consists of an on-the-fly creation of new mixtures from single speaker sources. In this work, we expanded this powerful technique by applying speed perturbation before mixing the sources. The speed randomly changes between 95 \% slow-down and 105 \% speed-up.
%We used randomly sampled independent speed changes for each source.

%(\dataaug=2 in Table \ref{tab:ablation}), and finally in addition to the speed change we also explored creating mixtures by randomly mixing the sources (\dataaug=3 in Table \ref{tab:ablation}) \cite{zeghidour2020wavesplit}. We refer to this option as dynamic mixing (DM) in Tables \ref{table:WSJ2Mix}, \ref{table:3mix}.  

We use the Adam algorithm \cite{kingma2017adam} as an optimizer, with a learning rate of $1.5e^{-4}$. After epoch 65 (after epoch 85 with \dataaug), the learning rate is annealed by halving it if we do not observe any improvement of the validation performance for 3 successive epochs (5 epochs for \dataaug). Gradient clipping is employed to limit the $l^2$-norm of the gradients to 5. During training, we used a batch size of 1, and used the scale-invariant signal-to-noise ratio (SI-SNR) \cite{le2019sdr} via utterance-level permutation invariant loss \cite{kolbaek2017multitalker}, with clipping at 30\,dB \cite{zeghidour2020wavesplit}. \cem{The SI-SNR measures the energy of the signal over the energy of the noise by using a zero mean signal estimate and the ground truth signal. We also use SDR (Signal to Distortion Ratio) which measures the energy of the signal over the energy of the noise, artifacts, and interference, as defined in \cite{vincent2006performance}. %Larger values of SI-SNR, and SDR indicates better separation performance. The values of both SI-SNR and SDR are continuous and are not bounded from below or above. 
In our tables, we report SI-SNRi, and SDRi values that denote the SI-SNR and SDR improvement values from the baseline results obtained when SI-SNR and SDR are computed with the mixture as the source estimate.}

We use automatic mixed precision to speed up training. Each model is trained for a maximum of 200 epochs. Each epoch takes approximately 1.5 hours on a single NVIDIA V100 GPU with 32 GB of memory with automatic mixed precision. The training recipes for SepFormer on WSJ0-2/3Mix datasets, Libri2/3Mix datasets, WHAM! and WHAMR! datasets can found in Speechbrain\footnote{\url{https://github.com/speechbrain/speechbrain}}. 

\begin{table}[t]
\centering
\caption{Best results on the WSJ0-2mix dataset (test-set). DM stands for dynamic mixing. } 
\vspace{0.2cm}
\label{table:WSJ2Mix}
\resizebox{9cm}{!}{
\begin{tabular}{l|c|c|c|c}
\textbf{Model} & \textbf{SI-SNRi} & \textbf{SDRi} & \textbf{\# Param} & \textbf{Stride } \\
\hline \hline 
Tasnet \cite{luo2017tasnet} & 10.8 & 11.1 & n.a. & 20  \\ \hline
SignPredictionNet \cite{wang2018deep} & 15.3 & 15.6 & 55.2M & 8\\ \hline
Conv-TasNet \cite{luo2018convtasnet} & 15.3  & 15.6 & 5.1M & 10\\ \hline 
Two-Step CTN \cite{tzinis2019twostep} & 16.1 & n.a. & 8.6M & 10 \\ \hline
MGST \cite{zhao2020MSGtransformer} & 17.0 & 17.3 & n.a. & n.a. \\ \hline
DeepCASA \cite{liu2019divide} & 17.7 & 18.0 & 12.8M & 1\\ \hline
FurcaNeXt \cite{shi2019furcanext} & n.a. & 18.4 & 51.4M & n.a.\\ \hline 
DualPathRNN \cite{luo2020dualpath} & 18.8 & 19.0 & 2.6M & 1 \\ \hline
sudo rm -rf \cite{tzinis2020sudo} &  18.9 & n.a. & 2.6M  & 10\\ \hline 
VSUNOS \cite{nachmani2020voice} & 20.1 & 20.4 & 7.5M & 2 \\ \hline
DPTNet* \cite{dptn} & 20.2 & 20.6 & 2.6M & 1\\  \hline 
\cem{DPTNet + DM**}  \cite{dptn} & 20.6 & 20.8 & 26.2M & 8\\  \hline 
\cem{DPRNN + DM**}  \cite{luo2020dualpath} & 21.8 & 22.0 & 27.5M & 8\\  \hline 
\cem{SepFormer-CF + DM**}  & 21.9 & 22.1 & 38.5M & 8\\  \hline 
Wavesplit*** \cite{zeghidour2020wavesplit} & 21.0 & 21.2 & 29M & 1 \\ 
Wavesplit*** + DM \cite{zeghidour2020wavesplit} & 22.2 & 22.3 & 29M & 1 \\ \hline
\hline
\textbf{SepFormer} & 20.4 & 20.5 & 25.7M & 8 \\ 
%\textbf{SepFormer + Speed} &  21.8 & 21.9 & 26M & 8 \\
\textbf{SepFormer + DM} &  \textbf{22.3} & \textbf{22.4} & 25.7M & 8 \\
\hline \hline
\end{tabular}
\vspace{1ex}
}
\vspace{1ex}
{\raggedright 
\footnotesize{*only SI-SNR and SDR (without improvement) are reported.} \par}
{\raggedright \footnotesize{**our experimentation} \par}
{\raggedright \footnotesize{***uses speaker-ids as additional info.} \par}
\end{table}

\usetikzlibrary{fit,shapes.misc}
\newcommand\marktopleft[1]{%
    \tikz[overlay,remember picture] 
        \node (marker-#1-a) at (0,1.5ex) {};%
}
\newcommand\markbottomright[1]{%
    \tikz[overlay,remember picture] 
        \node (marker-#1-b) at (0,0) {};%
    \tikz[overlay,remember picture,thick,dashed,inner sep=3pt]
        \node[draw,rounded rectangle,fit=(marker-#1-a.center) (marker-#1-b.center)] {};%
}

\begin{table}[t!]
    \caption{Ablation of the SepFormer on WSJ0-2Mix (validation set).}
    \vspace{0.1cm}
    \resizebox{8.6cm}{!}{
    \begin{tabular}{c|c|c | c| c | c | c | c  }
        \textbf{SI-SNRi} & \textbf{$N$} & $\numintra$ & $\numinter$ & \textbf{ \# Heads} & \textbf{DFF} &  \textbf{PosEnc} & \textbf{\dataaug}  \\ \hline \hline
        {22.3} & {2} & {8} & {8} & {8} & {1024} & {Yes} & \marktopleft{a6} Yes \\ \hline 
        %{21.8} & {2} & {8} & {8} & {8} & {1024} & {Yes} & 2 \\ \hline 
        %{21.2} & {2} & {8} & {8} & {8} & {1024} & {Yes} & 1 \\ \hline 
        {20.5} & {2} & \marktopleft{a4}{8} & {8} & {8} & \marktopleft{a5}{1024} & {Yes} & No \markbottomright{a6} \\ \hline 
        %{20.4}  & {2} & {4} & {4}  & {16} & {2048}&  {Yes} & No \\ \hline  % Mirco's best model
        %20.2 & 2 & 4 & 4  & 8 &2048 & \marktopleft{a3} Yes &  No  \\ \hline % dpt 35 (cedar) 
        %20.2 & 2 & 8 & 8 & 8 & 1024 & Yes \\ \hline  % Jian -- this got to 20.5!
        19.9 & 2 & 4 & 4 \markbottomright{a4}& 8 & 2048 \markbottomright{a5} & \marktopleft{a3}Yes & No \\ \hline % dpt 25
        %19.8 & \marktopleft{a4} 3 & 4 & 4  & 8 & 2048 & \marktopleft{a3}Yes & No \\ \hline  %dpt 26
        19.4 & 2 & 4 & 4 & 8 & 2048 & No \markbottomright{a3} & No \\ \hline  %dpt 18
        19.2 & 2 &\marktopleft{a2} 4 & 1 & 8 & 2048 & Yes & No\\ \hline %Mirko ablation
        18.3 & 2 & 1 & 4 \markbottomright{a2} & 8 & 2048 & Yes & No\\ \hline %Mirko ablation
        19.1  & 2 & 3 & 3  & 8 &2048 &  \marktopleft{a1} Yes & No \\ \hline % dpt22
        19.0  & 2 & 3 & 3  & 8  & 2048 & No \markbottomright{a1}  & No \\ \hline % dpt15
    \end{tabular}
    }
     \label{tab:ablation}
\end{table}

%\vspace{-0.3cm}

\section{Results}
\label{sec:experiments}

\subsection{Results on WSJ0-2/3 Mix datasets}
\label{sec:results-wsj0}

WSJ0-2/3 Mix datasets are standard benchmarks in the source-separation literature. In Table \ref{table:WSJ2Mix}, we compare the performance achieved by the proposed SepFormer with the best results reported in the literature on the WSJ0-2mix dataset. The SepFormer achieves an SI-SNR improvement (SI-SNRi) of 22.3\,dB and a Signal-to-Distortion Ratio \cite{vincent2006performance} (SDRi) improvement of 22.4\,dB on the test-set with dynamic mixing. When using dynamic mixing, the proposed architecture achieves state-of-the-art performance. The SepFormer outperforms previous systems without dynamic mixing, except for Wavesplit, which leverages speaker identity as additional information. 

In Table \ref{tab:ablation}, we also study the effect of various hyperparameters and data augmentations strategies (the reported performance is computed on the validation set).
We observe that the number of InterT and IntraT blocks has an important impact on the performance. The best performance is reached with 8 layers for both blocks replicated two times. We also would like to point out that a respectable performance of 19.2\,dB is obtained even when we use a single-layer transformer for the InterTransformer. \cem{In contrast, when a single layer Transformer is used for IntraTransformer the performance drops to 18.3\,dB}. This suggests that the IntraTransformer, and thus local processing, has a greater influence on the final performance. It also emerges that positional encoding is helpful (e.g. see lines 3-4, and 7-8 of Table \ref{tab:ablation}). A similar outcome has been observed in \cite{kim2020t} for speech enhancement. As for the number of attention heads, we observe a slight performance difference between 8 and 16 heads. Finally, as expected, we observed that dynamic mixing improves the performance.

\cem{To ascertain that SepFormer has better performance because of architectural choices (and not just because it has more parameters or dynamic mixing), we have compared the SepFormer with DPTNet, DPRNN, and with a SepFormer that utilizes a Conformer \cite{gulati2020conformer} block in the IntraTransformer. These three models are trained under the same conditions (with dynamic mixing, and $ kernelsize=16$, $kernelstride=8$ as SepFormer). The results of this experiment in terms of SI-SNR on the test are as follows (shown in Table \ref{table:WSJ2Mix}) 20.6\,dB for DPTNet (26.2M parameters), 21.8\,dB for DPRNN (27.5M parameters), and 21.9\,dB for SepFormer with Conformer block as intra block (38.5M parameters) (SepFormer-CF in Table \ref{table:WSJ2Mix}). We, therefore, observe that the better performance of SepFormer is due to the architectural differences, and not just because it has more parameters or is trained with dynamic mixing.}

Table \ref{table:3mix} showcases the best-performing models on the WSJ0-3mix dataset. SepFormer obtains state-of-the-art performance with an SI-SNRi of 19.5\,dB and an SDRi of 19.7\,dB.
We here used the best architecture found for the WSJ0-2mix dataset in Table \ref{tab:ablation}. The only difference is that the decoder has three outputs now.
It is worth noting that the SepFormer outperforms all previously proposed systems on this corpus. %In particular, the SepFormer outperforms Wavesplit \cite{zeghidour2020wavesplit} even though our model does not use speaker identities. 

\cem{Our results on WSJ0-2mix and WSJ0-3mix show that we can achieve state-of-the-art performance in separation with an RNN-free Transformer-based model. The major advantage of SepFormer over RNN-based systems like \cite{luo2020dualpath, nachmani2020voice, dptn} is the possibility to parallelize the computations over different time steps. This feature leads to faster training and inference, as described in the following section.}

 \begin{table}[t]
 \caption{Best results on the WSJ0-3mix dataset.}
 \vspace{0.1cm}
 \label{table:3mix}
 \centering
 \resizebox{8.6cm}{!}{
\begin{tabular}{l|c|c|c}
\textbf{Model} & \textbf{SI-SNRi} & \textbf{SDRi} & \textbf{\# Param} \\
\hline \hline 
Conv-TasNet \cite{luo2018convtasnet} & 12.7 & 13.1 & 5.1M\\ \hline 
DualPathRNN \cite{luo2020dualpath} & 14.7 & n.a & 2.6M \\ \hline
VSUNOS \cite{nachmani2020voice} & 16.9 & n.a & 7.5M  \\ \hline
Wavesplit \cite{zeghidour2020wavesplit} & 17.3 & 17.6 & 29M \\ 
Wavesplit \cite{zeghidour2020wavesplit} + DM & 17.8 & 18.1 & 29M \\ \hline \hline
\textbf{SepFormer} &  17.6 & 17.9 & 26M \\   
\textbf{SepFormer + DM} &  \textbf{19.5} & \textbf{19.7} & 26M \\ \hline  
\end{tabular}
}
\end{table}

\subsection{Speed and Memory Comparison}
\label{sec:speed}

\begin{figure*}[t!]
    \centering
    \includegraphics[width=0.292\textwidth]{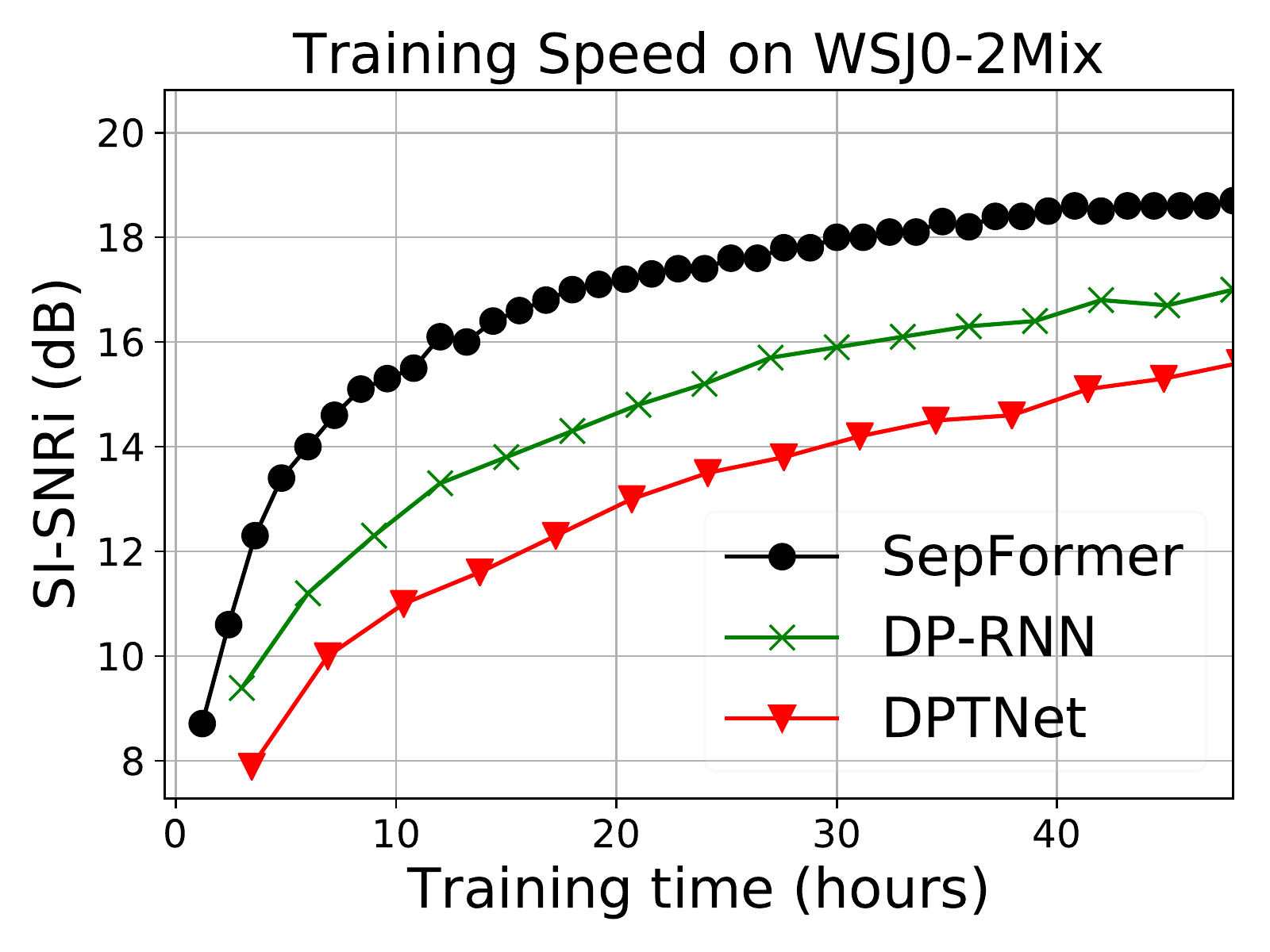}
    \includegraphics[width=.64\textwidth, trim=0cm 0cm 0cm 0cm, clip]{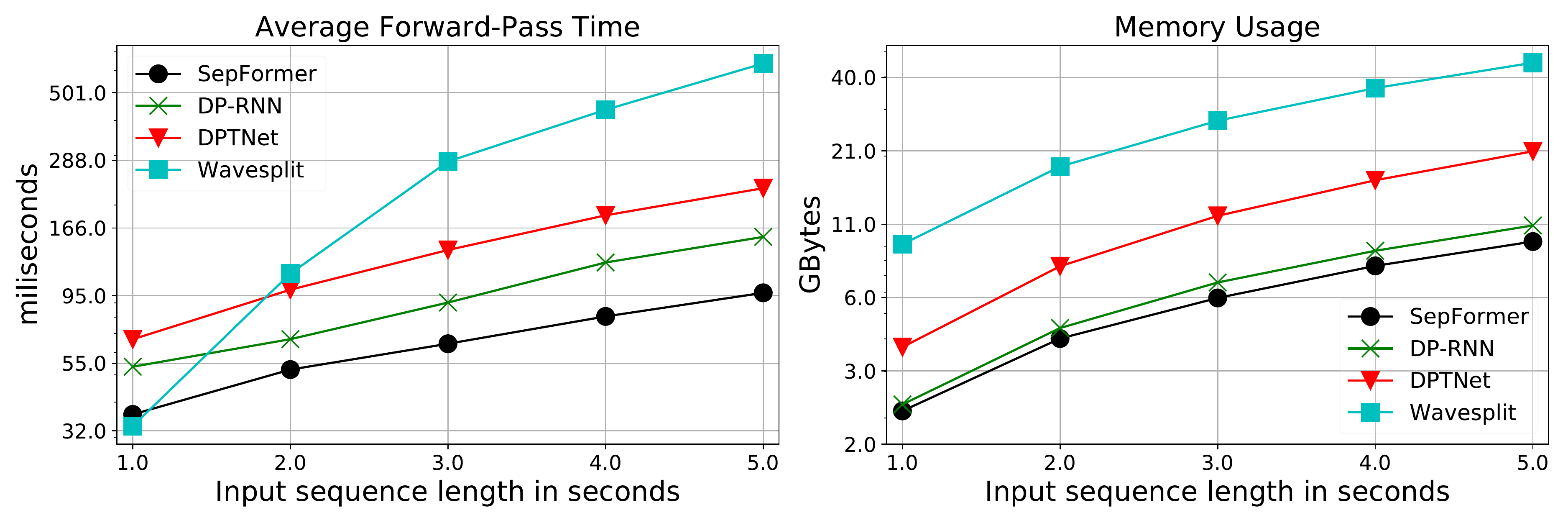}
    \vspace{-0.2cm}
    \caption{(\textbf{Left}) The training curves of SepFormer, DPRNN, and DPTNeT on the WSJ0-2mix dataset. \textbf{(Middle \& Right)} The comparison of forward-pass speed and memory usage in the GPU %for SepFormer, DPRNN, and DPTNeT 
    on inputs ranging 1-5 seconds long sampled at 8kHz.}
    \label{fig:forwardpass}
    \vspace{-0.2cm}
\end{figure*}

\begin{figure}
    \centering
    \includegraphics[width=.45\textwidth]{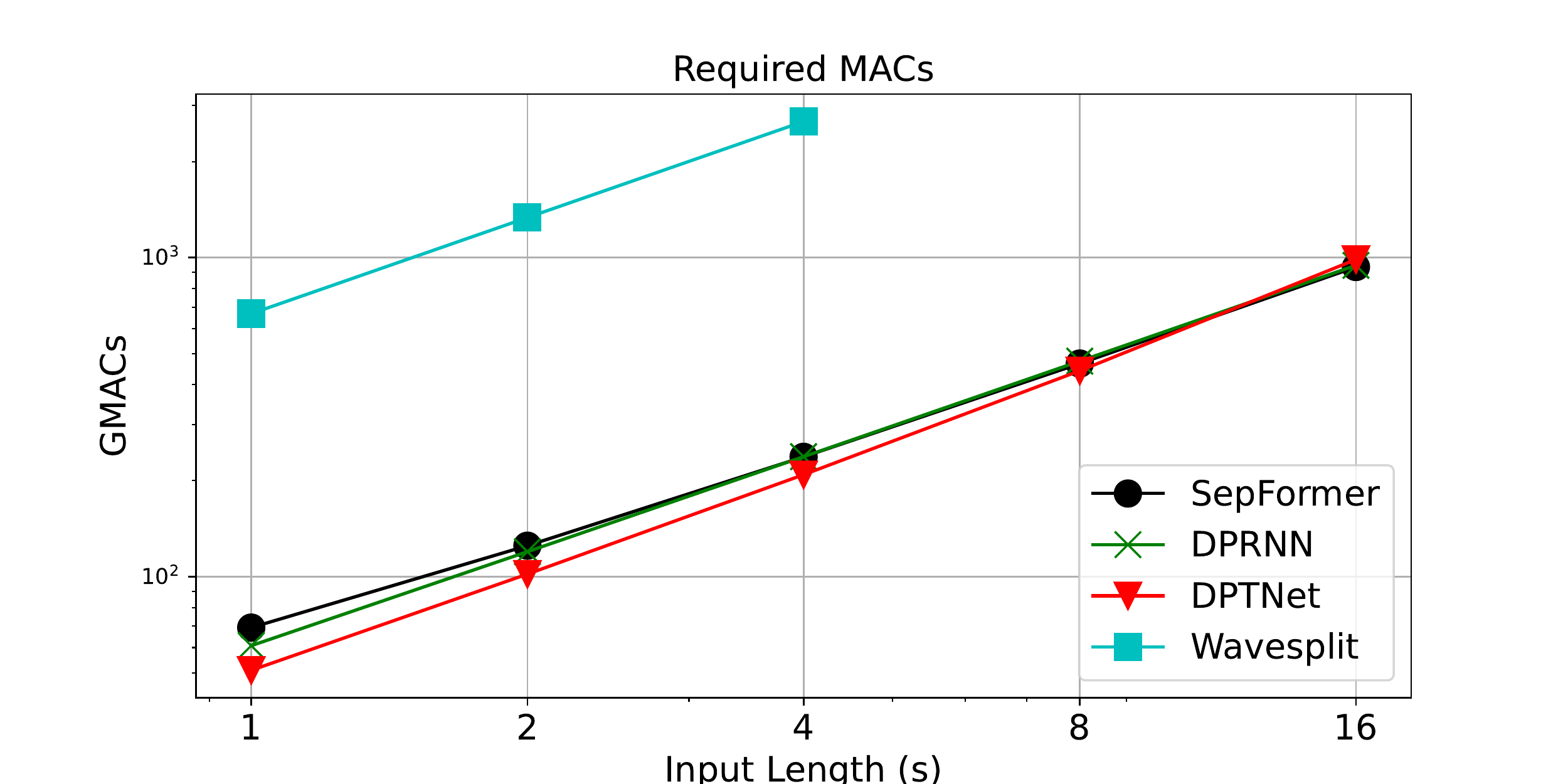}
    \caption{Comparing the required Multiply Accumulate Operations (MACs) for SepFormer, DPRNN, DPTNet and Wavesplit. }
    \label{fig:macsfirstfigure}
\end{figure}

We now compare the training and inference speed of our model with DPRNN \cite{luo2020dualpath} and DPTNet \cite{dptn}. Figure \ref{fig:forwardpass} (left) shows the performance achieved on the validation set in the first 48 hours of training versus the wall-clock time (on the WSJ0-2mix dataset).
For a fair comparison, we used the same machine with the same GPU (a single NVIDIA V100-32GB) for all the models. Moreover, all the systems are trained with a batch size of 1 and employ automatic mixed precision.  We observe that the SepFormer is faster than DPRNN and DPTNeT.  Figure \ref{fig:forwardpass} (left), highlights that SepFormer 
%our model (which achieves a final performance of 20.4 dB on the validation set of WSJ0-2Mix) 
reaches above 17\,dB levels only after a full day of training, whereas the DPRNN model requires two days of training to achieve the same level of performance.  

Figure \ref{fig:forwardpass} (middle\&right) compares the average computation time (in ms) and the total memory allocation (in GB) during inference when single precision is used. We analyze the speed of our best model for both WSJ0-2Mix and WSJ0-3Mix datasets. We compare our models against DP-RNN, DPTNet, and Wavesplit. All the models run in the same NVIDIA RTX8000-48GB GPU using the PyTorch profiler \cite{pytorch-profiler}. 

From this analysis, it emerges that the SepFormer is faster and less memory-demanding than DPTNet, DPRNN, and Wavesplit. We observed the same behavior using the CPU for inference. Such a level of computational efficiency is achieved even though the proposed SepFormer employs more parameters than the other RNN-based methods (see Table \ref{table:WSJ2Mix}).
This is not only due to the superior parallelization capabilities of the proposed model, but also because the best performance is achieved with a stride factor of 8 samples, against a stride of 1 for DPRNN and DPTNet. Increasing the stride of the encoder results in downsampling the input sequence, and therefore the model processes fewer data. In \cite{luo2020dualpath}, the authors showed that the DPRNN performance degrades significantly when increasing the stride factor. %, although this results in a faster model. 
The  SepFormer, instead, reaches competitive results even with a relatively large stride, leading to the aforementioned speed and memory advantages. \cem{We also compare the required Multiply-Accumulate Operations (MACs) for SepFormer, DPRNN, DPTNet, and Wavesplit in Figure \ref{fig:macsfirstfigure}. We see that SepFormer, DPRNN, and DPTNet have roughly the same MACs requirements. Despite that, due to the better parallelizability of SepFormer, the inference time (forward-pass time) is faster for SepFormer (and less memory demanding).
We also observe that Wavesplit has a significantly larger requirement in terms of MACs. We could not obtain MACs for sequences longer than 4 seconds due to out-of-memory errors for Wavesplit.}

\subsection{Results on WHAM! and WHAMR! datasets}
WHAM! \cite{wichern2019wham} and WHAMR! \cite{maciejewski2020whamr} datasets are versions of the WSJ0-2Mix dataset with noise and  noise+reverberation, respectively. We use the same model configuration as the WSJ0-2/3Mix dataset for WHAM! and WHAMR!. For both datasets, we train the model so that it does source separation and speech enhancement at the same time. With the WHAM! dataset the model learns to denoise while also separating. With WHAMR!, the model jointly learns to denoise and dereverberation in addition to separating. In this case, we augment the mixtures by randomly choosing a room-impulse response from the training set of the WHAMR! dataset.

In Tables \ref{table:wham}, \ref{table:whamr}, we provide results with SepFormer on the WHAM! and WHAMR! datasets compared to the other methods in the literature. We observe that on both WHAM! and WHAMR! datasets SepFormer outperforms the previously proposed methods even without using DM, which further improves the performance.

\begin{table}[t]
\centering
\small
 \caption{Best results on the WHAM dataset.}
 \vspace{0.2cm}
 \label{table:wham}
 \centering
 \resizebox{7.6cm}{!}{
\begin{tabular}{l|c|c}
\textbf{Model} & \textbf{SI-SNRi} & \textbf{SDRi}  \\
\hline \hline 
Conv-TasNet  & 12.7 & - \\ \hline 
Learnable fbank & 12.9 & - \\ \hline 
MGST \cite{zhao2020MSGtransformer} & 13.1 & - \\ \hline
Wavesplit + DM \cite{zeghidour2020wavesplit} & 16.0 & 16.5  \\ \hline \hline
\textbf{SepFormer} & 14.7 & 15.1  \\ 
\textbf{SepFormer + SpeedA.} & 16.3 & 16.7  \\ 
\textbf{SepFormer + DM} & \textbf{16.4} &  \textbf{16.7} \\ \hline
\end{tabular}
}
\end{table}

\begin{table}[t]
\small
 \caption{Best results on the WHAMR dataset.}
 \vspace{0.1cm}
 \label{table:whamr}
 \centering
 \resizebox{7.6cm}{!}{
\begin{tabular}{l|c|c}
\textbf{Model} & \textbf{SI-SNRi} & \textbf{SDRi}  \\
\hline \hline 
Conv-TasNet  & 8.3 & - \\ \hline 
BiLSTM Tasnet & 9.2 & - \\ \hline 
Wavesplit  + DM \cite{zeghidour2020wavesplit}& 13.2 & 12.2  \\ \hline \hline
\textbf{SepFormer} & 11.4 &  10.4 \\
\textbf{SepFormer + SpeedA.} &  13.7 & 12.7  \\   
\textbf{SepFormer + DM} & \textbf{14.0} & \textbf{13.0} \\ \hline  
\end{tabular}
}
\end{table}

\subsection{Results on Libri-2/3Mix datasets}

LibriMix \cite{cosentino2020librimix} is proposed as an alternative to WSJ0-2/3Mix with more diverse and longer utterances.
Table \ref{table:librimix} shows results achieved with the Libri-2/3Mix datasets (both for cleaning any noisy versions). We use here the same model configuration found on WSJ0-2mix in Table \ref{tab:ablation}. Similar to the WHAM! dataset, for Libri-2/3 Mix datasets the network is trained to separate and denoise at the same time. We train the models on the train-360 set, both with and without dynamic mixing. In addition to providing the performance of the models trained on LibriMix, we also report the performance of the model pre-trained on the WSJ0-Mix dataset. One common criticism of the WSJ0-Mix is that the models trained on it do not generalize well to other tasks. Here, we showcase that SepFormer can obtain respectable performance on Libri 2/3 Mix clean versions even when trained on WSJ0-2/3Mix, surpassing the performance of a Conv-TasNet model trained on LibriMix as shown in row \textbf{SepFormer PT} of Table \ref{table:librimix}. We also report the performance obtained with a model pretrained on WHAM! dataset. 

When trained directly on Libri2/3-Mix with dynamic mixing, SepFormer obtains the state-of-the-art performance as shown in row \textbf{SepFormer + DM} of Table \ref{table:librimix}. We also show the case where we fine-tune a pretrained SepFormer model on LibriMix in the row \textbf{SepFormer + DM PT+FT} of Table \ref{table:librimix}. 

\begin{table*}[t]
 \caption{Results on Libri2Mix and Libri3Mix datasets. SepFormer + PT indicates pretraining on WSJ0-2/3 Mix or WHAM! dataset according to the case. SepFormer + DM PT+FT is the case where the pretrained SepFormer model is fine-tuned on LibriMix.}
 \vspace{0.1cm}
 \label{table:librimix}
 \centering
 \resizebox{17.0cm}{!}{
\begin{tabular}{l|c|c|c|c|c|c|c|c|}
 &  \multicolumn{2}{|c|}{Libri2Mix-Clean} & \multicolumn{2}{|c|}{Libri2Mix-Noisy} &
 \multicolumn{2}{|c|}{Libri3Mix-Clean} &
 \multicolumn{2}{|c|}{Libri3Mix-Noisy} \\ \hline \hline
\textbf{Model} & \textbf{SI-SNRi} & \textbf{SDRi} &  \textbf{SI-SNRi} & \textbf{SDRi} &
\textbf{SI-SNRi} & \textbf{SDRi} & 
 \textbf{SI-SNRi} & \textbf{SDRi} \\
\hline \hline 
Conv-TasNet  & 14.7 & - & 12.0 & - & 10.4 & - & 10.4& -   \\ \hline 
\textbf{SepFormer PT} 
&  17.0 & 17.5 & 11.2 & 13.1 &15.0 & 15.6 & n.a. & n.a. \\ \hline  
Wavesplit \cite{zeghidour2020wavesplit}
&19.5 & 20.0 & 15.1 & 15.8 & 15.8 & 16.3 & 13.1 & 13.8  \\ \hline
Wavesplit + DM \cite{zeghidour2020wavesplit}
& 20.5 & 20.0 & 15.2 & 15.9 &17.5 & 18.0 & 13.4 & 14.1  \\ \hline \hline
\textbf{SepFormer} 
& 19.2 & 19.4  & 14.9  & 15.4  & 16.9 & 17.3  & 14.3  & 14.8    \\  
\textbf{SepFormer + DM} 
& 20.2 & 20.5 & 15.9  & 16.5  & 18.2 & 18.6 & \textbf{15.0}  & \textbf{15.5}  \\  
%\textbf{SepFormer SpeedAug PT+FT}  
%&  20.1  & 20.3 & & & 18.4 & 18.6 & &  \\
\textbf{SepFormer + DM PT+FT} 
& \textbf{20.6} & \textbf{20.8} & \textbf{15.9} & \textbf{16.5} & \textbf{19.0} & \textbf{19.4} & n.a. & n.a. \\ \hline
\end{tabular}
}
\end{table*}

%\begin{table}[h!]
% \caption{Libri3Mix.}
% \vspace{0.1cm}
% \label{table:3mix}
% \centering
% \resizebox{7.6cm}{!}{
%\begin{tabular}{l|c|c}
%\textbf{Model} & \textbf{SI-SNRi} & \textbf{SDRi}  \\
%\hline \hline 
%Conv-TasNet  & 10.4 & - \\ \hline 
%\textbf{SepFormer trained on WSJ0-3Mix} &  15.0 & 15.6  \\ \hline  
%Wavesplit & 15.8 & 16.3 \\ \hline
%Wavesplit  + DM & 17.5 & 18.0  \\ \hline \hline
%\textbf{SepFormer SpeedAug} & 18.4 & 18.6 \\
%\textbf{SepFormer DM} & 18.7 & 19.0 \\ \hline
%\end{tabular}
%}
%\end{table}

\begin{figure}[t]
    \centering
    \includegraphics[width=0.51\textwidth, trim=1.5cm 0cm 1.5cm 0cm, clip]{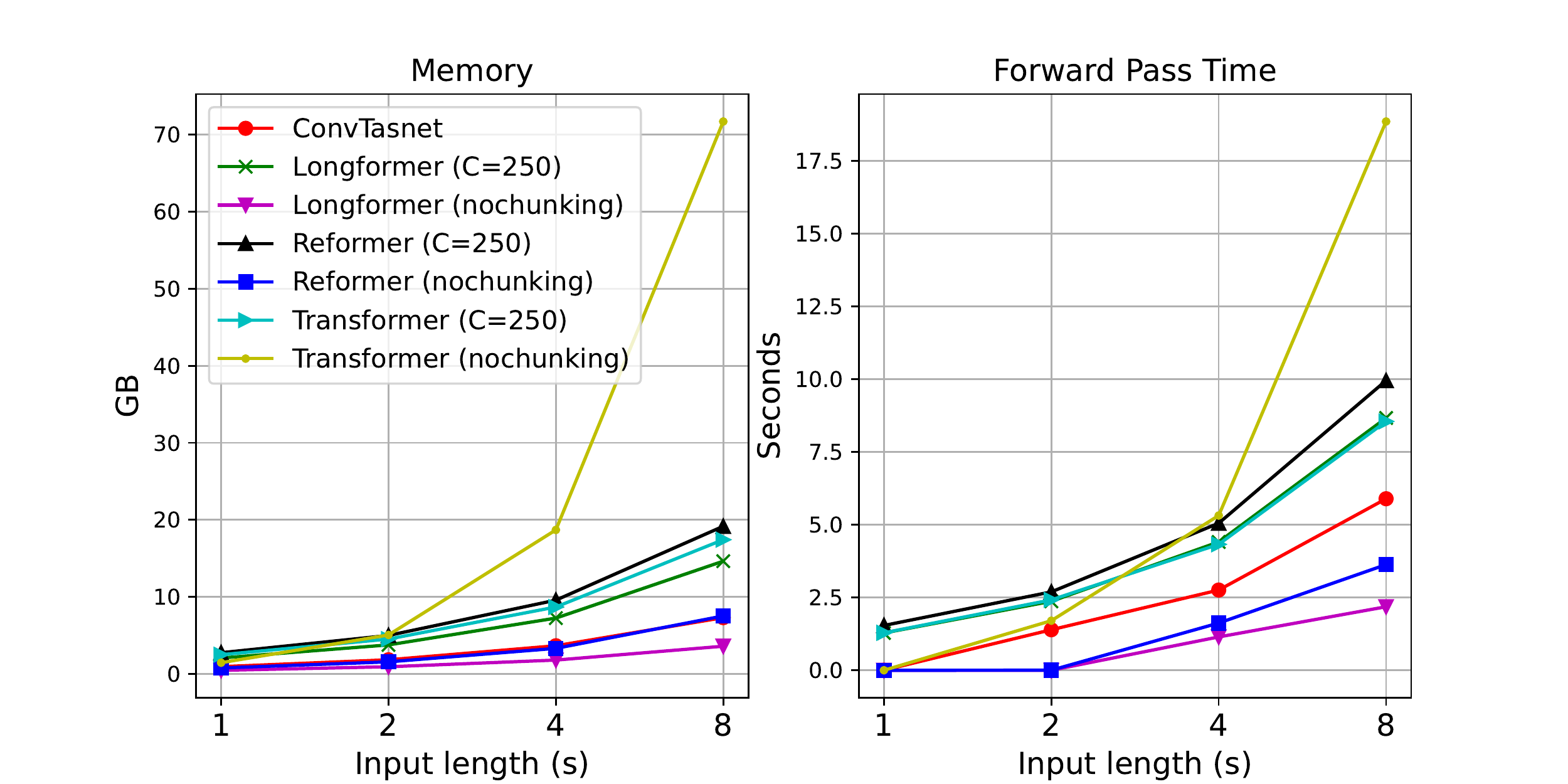}
    \caption{Comparison of memory usage and forward-pass time on the SepFormer architecture with several different self-attention types.}
    \label{fig:efficiency}
\end{figure}

\subsection{Ablations on the type of Self-Attention}
%In order to be able to fairly assess the performance of SepFormer compared to the other efficient transformer architectures, we try the efficient transformer architectures within the dual-path processing pipeline, and just on its own. More specifically, 
%In this section, we compare the performance and computational advantages of the different self-attention mechanisms we described in Section \ref{sec:efficient-self-attention}. In detail, i

\begin{table*}[t!]
 \caption{Comparison of different types of self-attention (WSJ0-2Mix, test set). %The column $\chnksize=250v1$ indicates the case where chunk size is 250 and the IntraT/InterT use the attention mechanism indicated in the first column. The column $\chnksize=250v2$ indicates the case where chunk size is 250, IntraT uses the attention mechanism indicated in the first column, and InterT uses the full attention mechanism. The column $\chnksize=1000$ indicates the case where the chunk size is 1000 and IntraT uses the attention mechanism indicated in the first column. The column No Chunking indicates the case where we do not have an InterT block.
 }
 \vspace{0.1cm}
 \label{table:attentiontype}
 \centering
\resizebox{17.6cm}{!}{
\begin{tabular}{l|c|c|c|c|c|c|c|c|}
 & \multicolumn{2}{|c|}{\chnksize=250v1} & \multicolumn{2}{|c|}{\chnksize=250v2} & \multicolumn{2}{|c|}{\chnksize=1000} & \multicolumn{2}{|c|}{No Chunking} 
  \\ \hline 
\textbf{Model} & \textbf{SI-SNRi} & \textbf{SDRi} & \textbf{SI-SNRi} & \textbf{SDRi} & \textbf{SI-SNRi} & \textbf{SDRi} & \textbf{SI-SNRi} & \textbf{SDRi} \\
\hline \hline 
Longformer & 19.44 & 19.66  & 19.32 & 19.54 & 18.17 & 18.39 & 13.11  & 13.36 \\ \hline 
%Longformer with $\numchnks=1000$ & 16.5 & - \\ \hline 
Linformer & 2.75 & 3.01 & 6.04 & 6.30 & 4.74 & 5.02 & 6.39 & 6.68 \\ \hline 
%Linformer with $\numchnks=1000$ &  & - \\ \hline 
Reformer & 20.09 & 20.28 & 20.34 & 20.52 & 19.42 & 19.61 & 16.66 & 16.86 \\ \hline 
%Reformer with $\numchnks=1000$ &  & - \\ \hline 
\textbf{Transformer (SepFormer)} & \textbf{21.61} & \textbf{21.79}  & 21.61 & 21.79 & 21.48 & 21.66 & 12.40 & 12.60 \\ \hline 
\end{tabular}
}
\end{table*}

Table \ref{table:attentiontype} compares the WSJ0-2Mix test set performance of four different architectural choices for Full (regular), LongFormer, Linformer, and Reformer self-attention. We train all models using only speed augment (no dynamic mixing), using 4-second long training segments. The architectural choices in this ablation experiment given in Table \ref{table:attentiontype} are summarized as follows:

\begin{itemize}
    \item (second column, \chnksize=250v1) in the IntraTransformer and InterTransformer blocks, we use the attention indicated in the first column. The chunk size for the IntraTransformer is set to $\chnksize=250$. 
    \item (third column, \chnksize=250v2) in the IntraTransformer we use the self-attention mechanism indicated in the first column, and in the InterTransformer we use the full attention mechanism. The chunk size for the IntraTransformer is set to $\chnksize=250$. 
    \item (fourth column, \chnksize=1000) in the IntraTransformer we use the self-attention mechanism indicated in the first column, and in the InterTransformer we use the full attention mechanism. The chunk size for the IntraTransformer is set to $\chnksize=1000$.  
    \item (fifth column, No Chunking) we do not apply chunking, and therefore only the IntraTransformer block is applied. 
    %We have observed out-of-memory (O.O.M.) behavior on the test-set for the full self-attention due to long sequences, and therefore we denote the corresponding entry by O.O.M.. 
\end{itemize}

We observe from Table \ref{table:attentiontype} that the architectural choices adopted in the original SepFormer paper lead to the best overall performance in terms of SI-SNR and SDR improvement. We also show the memory usage and forward pass time (in CPU using Pytorch profiler \cite{pytorch-profiler}) for some of the relevant models in Figure \ref{fig:efficiency}. The chunking/dual-path mechanism in the SepFormer architecture reduces the memory footprint significantly compared to the architecture where no chunking is applied. From Figure \ref{fig:efficiency}, we also see that using the Longformer or Reformer blocks on the whole sequence without applying chunking is an efficient alternative to dual-path processing. Furthermore from Figure \ref{fig:efficiency}, we observe that compared to Conv-TasNet, Reformer without chunking results in almost equivalent memory usage, and faster forward pass time, while yielding slightly better performance, namely 16.7\,dB SI-SNRi on the test set of WSJ0-2Mix, compared to 15.3\,dB SI-SNRi obtained with Conv-TasNet \cite{luo2018convtasnet}.

We also observe that full attention is more efficient than using a reformer inside the dual-path pipeline in terms of forwarding pass time and memory usage, and full attention is not significantly more expensive than Longformer in terms of memory. Another observation is the fact that Linformer does not yield competitive performance. We suspect that this stems from the fact that Linformer has a reduced modeling capacity since it projects the time dimension inside the attention mechanism to a lower dimensionality, thereby losing time resolution, which is essential for effective source separation.

\begin{figure}[t]
    \centering
    \includegraphics[width=0.5\textwidth]{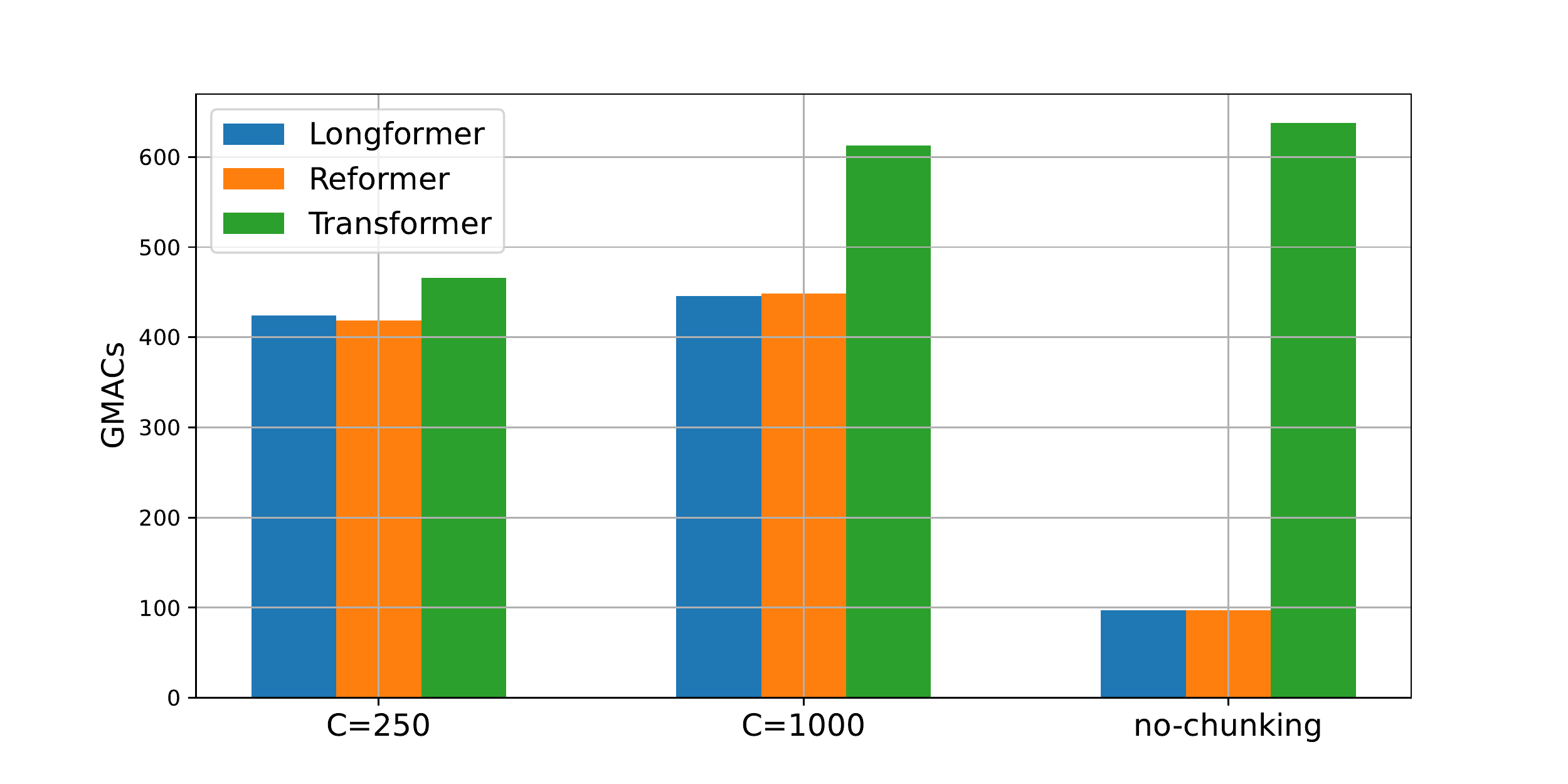}
    \caption{Multiply-Accumulate Operations for $\chnksize=250v1$, $\chnksize=1000$, and no chunking, obtained on a 4 seconds input signal. Reformer and Longformer provide significant reduction in MACs for the no-chunking option.}
    \label{fig:macs}
\end{figure}

\cem{To showcase the reduction of computational requirements when using efficient self-attention mechanisms (namely Longformer and Reformer), we report the total MACs (Multiply-Accumulate Operations) for a 4-seconds input sequence in Figure \ref{fig:macs}. We notice a significant reduction in the required MACs for Longformer and Reformer compared to the full-attention case. \cemAQ{In the case where no-chunking is applied,} from Table \ref{table:attentiontype} we also observe that the performance is significantly better with Reformer compared to the full-attention case. However, another important conclusion to make from this ablation study is that even though the Longformer and Reformer provide a reduction in computational requirements the best performance is obtained in the SepFormer architecture (that is the case where we use full-self attention in Intra and Inter transformer blocks).}

\subsection{Speech Enhancement}
We trained the SepFormer for speech enhancement on the WHAM! and WHAMR! datasets for noisy and noisy-reverberant mixtures with one speaker. The obtained enhancement performances in terms of SI-SNR, SDR, and PESQ \cite{Pesq} are presented in Table \ref{table:enhancement-wham}, and Table \ref{table:enhancement-whamr} respectively. 
Note that in WHAM! dataset the model is trained to denoise and on the WHAMR! dataset the model is trained to both dereverberation and denoising at the same time. The SepFormer models are trained to maximize the output SI-SNR between the estimated and the ground truth signals in the time domain. 

In addition to training  SepFormer, we also trained a Bidirectional LSTM, CNNTransformer, and 2D-CNN models from Speechbrain \cite{speechbrain}, which are trained to minimize the Euclidean distance of the estimated magnitude spectrogram and the spectrogram of the clean signals. We observe that SepFormer outperforms these methods by a large margin in terms of SI-SNR, SDR, and PESQ.   

\cem{Other than WHAM! and WHAMR! datasets, we have also trained a SepFormer on the Voicebank-DEMAND speech enhancement dataset for denoising. 
In this case,  the SepFormer estimates a mask applied on the magnitude of Short-time Fourier Transform (STFT) representation. %This is done because of the limited size of this dataset, which makes it challenging to learn effective filterbanks directly from the data. 
This allows a direct comparison with most state-of-the-art VoiceBank-DEMAND models, which also employ a mask on magnitude spectra, and compare directly the architectures. The results are given in Table \ref{table:enhancement-voicebank}. 
We observe that SepFormer outperforms most of the recent methods, except for the MetricGAN+ \cite{metricganplus}, and AIAT \cite{dualbranchyu2022}. Note that MetricGAN+ is optimized specifically on PESQ and, therefore, the performance in terms of the STOI metric is not reported. \cemAQ{We note that SepFormer obtains the same performance as AIAT (Magnitude version) in terms of the STOI metric \cite{STOI}.} These speech enhancement experiments, therefore, demonstrate that SepFormer is competitive also with respect to state-of-the-art denoising models.}

\begin{table}[t!]
\small
 \caption{Speech enhancement results on WHAM! dataset (denoising)}
 \vspace{0.1cm}
 \label{table:enhancement-wham}
 \centering
 \resizebox{7.6cm}{!}{
\begin{tabular}{l|c|c|c}
\textbf{Model} & \textbf{SI-SNR} & \textbf{SDR} & \textbf{PESQ}  \\
\hline \hline
{2D-CNN} & 7.90 & 8.70 & 2.36\\ \hline
{2D-CNN+BLSTM} & 7.09 & 7.89 & 2.32 \\ \hline
{BLSTM} & 5.15 & 6.2 & 2.11 \\ \hline
{CNNTransformer} & 8.3 & 8.9 & 2.52 \\ \hline
\textbf{SepFormer} & \textbf{14.35} & \textbf{15.04} & \textbf{3.07}   \\ \hline
\end{tabular}
}
\end{table}

\begin{table}[t!]
 \small
 \caption{Speech enhancement results on WHAMR! dataset (Denoising + Dereverberation)}
 \vspace{0.1cm}
 \label{table:enhancement-whamr}
 \centering
 \resizebox{7.6cm}{!}{
\begin{tabular}{l|c|c|c}
\textbf{Model} & \textbf{SI-SNR} & \textbf{SDR} & \textbf{PESQ}  \\
\hline \hline 
{2D-CNN} & 6.84 & 7.75 & 2.18 \\ \hline
{2D-CNN+BLSTM} & 5.87 &6.82  & 2.15 \\ \hline
{BLSTM} & 5.58 & 6.49 & 2.11 \\ \hline
{CNNTransformer} & 7.38 & 8.21 & 2.27 \\ \hline
\textbf{SepFormer} & \textbf{10.58} & \textbf{12.29} & \textbf{2.84} \\ \hline
\end{tabular}
}
\end{table}

\begin{table}[t!]
 \small
 \caption{Speech enhancement results on the Voicebank-Demand dataset (Denoising)}
 \vspace{0.1cm}
 \label{table:enhancement-voicebank}
 \centering
 \resizebox{7.0cm}{!}{
\begin{tabular}{l|c|c}
\textbf{Model} & \textbf{PESQ} & \textbf{STOI}  \\
\hline \hline 
{CNNTransformer} \cite{speechbrain} & 2.65  & 91.5 \\ \hline
{MetricGAN} \cite{fu2019metricGAN} & 2.86 & -   \\ \hline
{CRGAN} \cite{zhangcrgan} & 2.92 & 94.0  \\ \hline
{DeepMMSE} \cite{Hu2020DCCRNDC} & 2.95 & - \\
\hline
{YinPHASEN} \cite{yinphasen} & 2.99 & - \\
\hline
\textbf{MetricGAN+} \cite{metricganplus} & \textbf{3.15} & - \\
\hline
\textbf{AIAT (Magnitude)} \cite{dualbranchyu2022} & 3.11 & \textbf{94.9} \\
\hline 
\textbf{SepFormer} & 3.03 & \textbf{94.9} \\
\hline

\end{tabular}
}
\end{table}

\section{Conclusion}
In this paper, we studied in-depth Transformers for speech separation.  In particular, we built upon our previously proposed  SepFormer, an attention-only masking network based on dual-path processing. We extended our previous findings by performing additional experiments on more challenging and realistic datasets, such as LibriMix, WHAM!, WHAMR!. With all these datasets, the SepFormer achieves competitive or state-of-the-art performance.
We also analyzed the computational resources needed by popular separation models, and we showed that the SepFormer excels in terms of 
memory usage and forward-pass time given the state-of-the-art performance.
 As another contribution, we compared more efficient self-attention mechanisms such as Longformer, Linformer, and Reformer. 
Our results suggest that Longformer and especially Reformer are suitable for speech separation applications and obtain a highly favorable trade-off between performance and computational requirements.  Finally, we adapted the SepFormer to perform speech enhancement and showed competitive performance in this context as well.

%To foster replicability, the training recipes and pre-trained models for the SepFormer on WSJ0-2/3Mix, Libri2/3Mix, WHAM!, and WHAMR! are publicly available on Speechbrain toolkit \cite{speechbrain}.
%expanded upon our earlier conference paper and included evidence from more datasets for efficacy of SepFormer. We have also explored incorporating efficient self-attention mechanisms in the SepFormer architecture, and observed while the original SepFormer provides the best performance overall, employing the Reformer architecture without chunking results in a more efficient model while not sacrificing too much from performance. 

%
\bibliographystyle{IEEEtran}
\bibliography{refs}

\begin{wrapfigure}{l}{35mm} 
    \includegraphics[width=3.4cm,trim={4.2cm 0cm 2.4cm 0cm},clip,keepaspectratio]{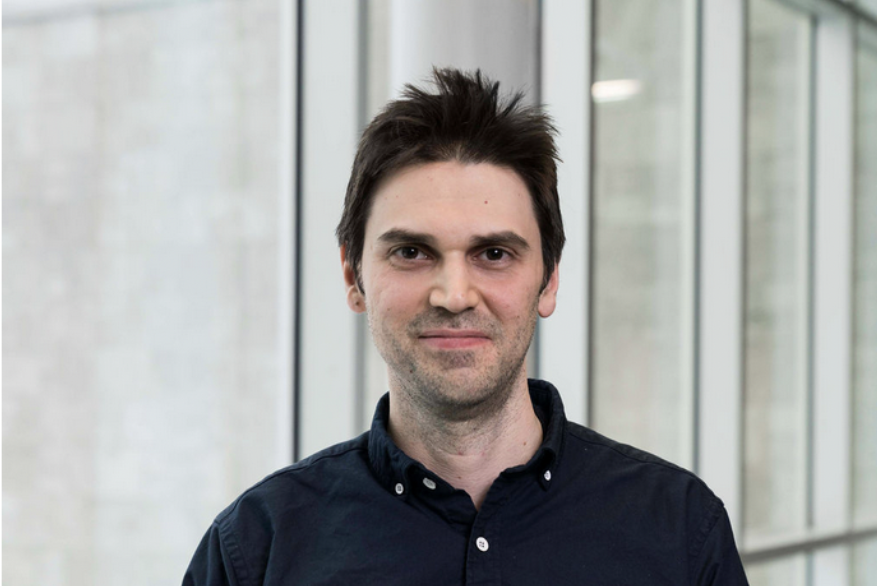}
\end{wrapfigure}\par

\noindent \textbf{Cem Subakan} is an Assistant Professor at Universit\'{e} Laval in the Computer Science and Software Engineering department. He is also currently an Affiliate Assistant Professor in the Concordia University Computer Science and Software Engineering Department, and an invited researcher at Mila-Québec AI Institue. He received his PhD in Computer Science from University of Illinois at Urbana-Champaign (UIUC), and did a postdoc in Mila Québec AI Institute and Université de Sherbrooke. He serves as reviewer in several conferences including NeurIPS, ICML, ICLR, ICASSP, MLSP and journals such as IEEE Signal Processing Letters (SPL), IEEE Transactions on Audio, Speech, and Language Processing (TASL). His research interests include Deep learning for Source Separation and Speech Enhancement under realistic conditions, Neural Network Interpretability, and Latent Variable Modeling. He is a recipient of best paper award in the 2017 version IEEE Machine Learning for Signal Processing Conference (MLSP), as well as the Sabura Muroga Fellowship from the UIUC CS department. He's a core contributor to the SpeechBrain project, leading the speech separation part. \\

\begin{wrapfigure}{l}{28mm} 
    \includegraphics[width=2.7cm,trim=0cm 0cm 0cm .5cm,clip,keepaspectratio]{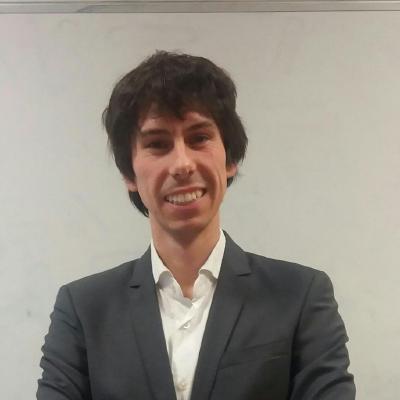}
\end{wrapfigure}\par

\noindent\textbf{Mirco Ravanelli} is an assistant professor at Concordia University, an adjunct professor at Université de Montréal, and Mila associate member. His main research interests are deep learning and Conversational AI. He is the author or co-author of more than 60 papers on these research topics. He received his Ph.D. (with cum laude distinction) from the University of Trento in December 2017. Mirco is an active member of the speech and machine learning communities. He is the founder and leader of the SpeechBrain project which aim to build an open-source toolkit for conversational AI and speech processing. \\

\begin{wrapfigure}{l}{28mm} 
    \includegraphics[width=2.7cm,trim=0cm 0cm 0cm 6.5cm,clip,keepaspectratio]{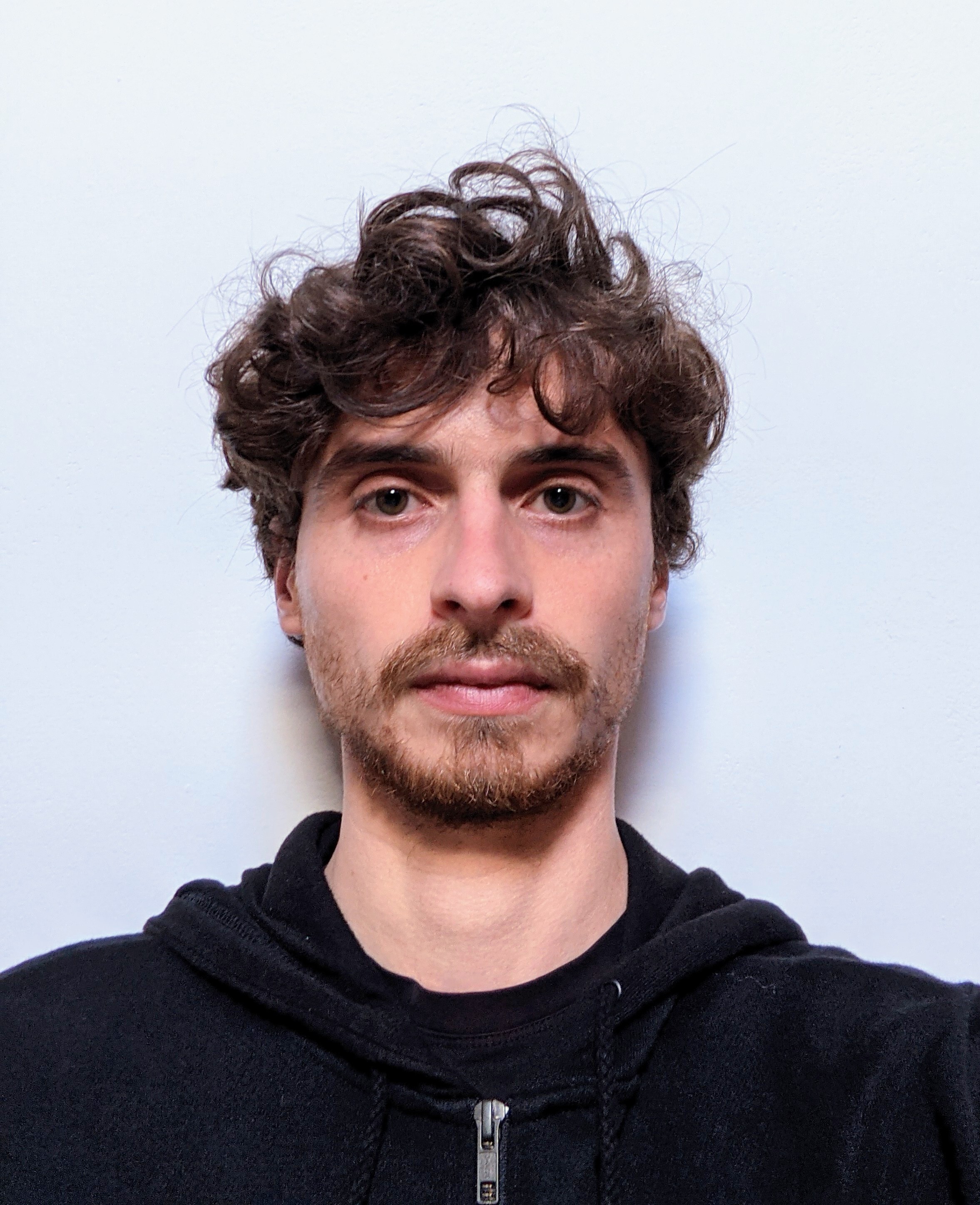}
\end{wrapfigure}\par

\noindent\textbf{Samuele Cornell} received a Master degree in Electronic Engineering at Università Politecnica delle Marche (UnivPM) in 2019. He is currently a doctoral candidate at UnivPM Department of Information Engineering. His current research interests are in the area of front-end pre-processing techniques for Automatic Speech Recognition applications such as: Source Separation, Speech Enhancement, Speech Segmentation and Beamforming. \\

\begin{wrapfigure}{l}{28mm} 
    \includegraphics[width=2.7cm,trim=0cm 0cm 0cm .0cm,clip,keepaspectratio]{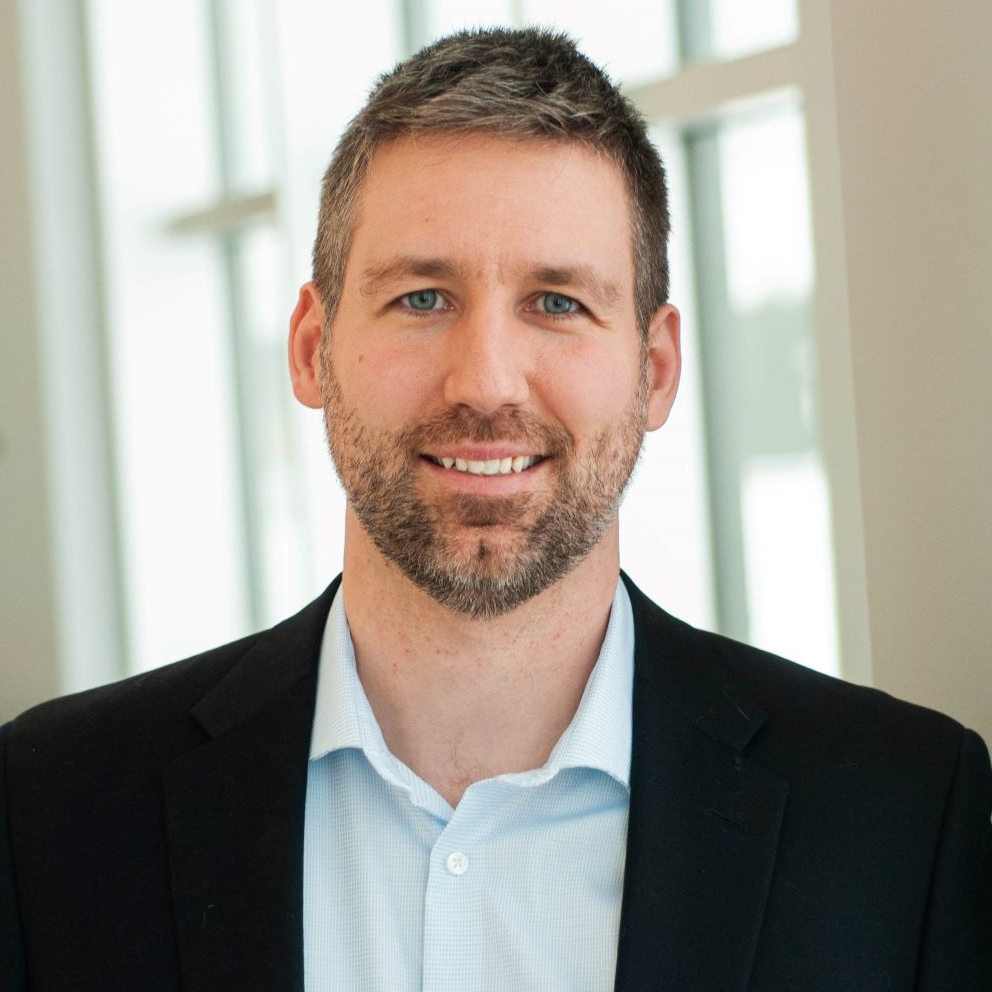}
\end{wrapfigure}\par

\noindent\textbf{François Grondin} received the B.Sc. degree in electrical engineering from McGill University, Montreal, QC, Canada, in 2009, and the M.Sc. and Ph.D. degrees in electrical engineering from the Université de Sherbrooke, Sherbrooke, QC, Canada, in 2011 and 2017, respectively. After completing postdoctoral work with the Computer Science and Artificial Intelligence Laboratory, Massachusetts Institute of Technology, Cambridge, MA, USA, in 2019, he became a Faculty Member with the Department of Electrical Engineering and Computer Engineering, Université de Sherbrooke. He is a member of the Ordre des ingénieurs du Québec. His research interests include robot audition, sound source localization, speech enhancement, sound classification, and machine learning. \\

\begin{wrapfigure}{l}{28mm} 
    \includegraphics[width=2.7cm,trim=0cm 0cm 0cm .0cm,clip,keepaspectratio]{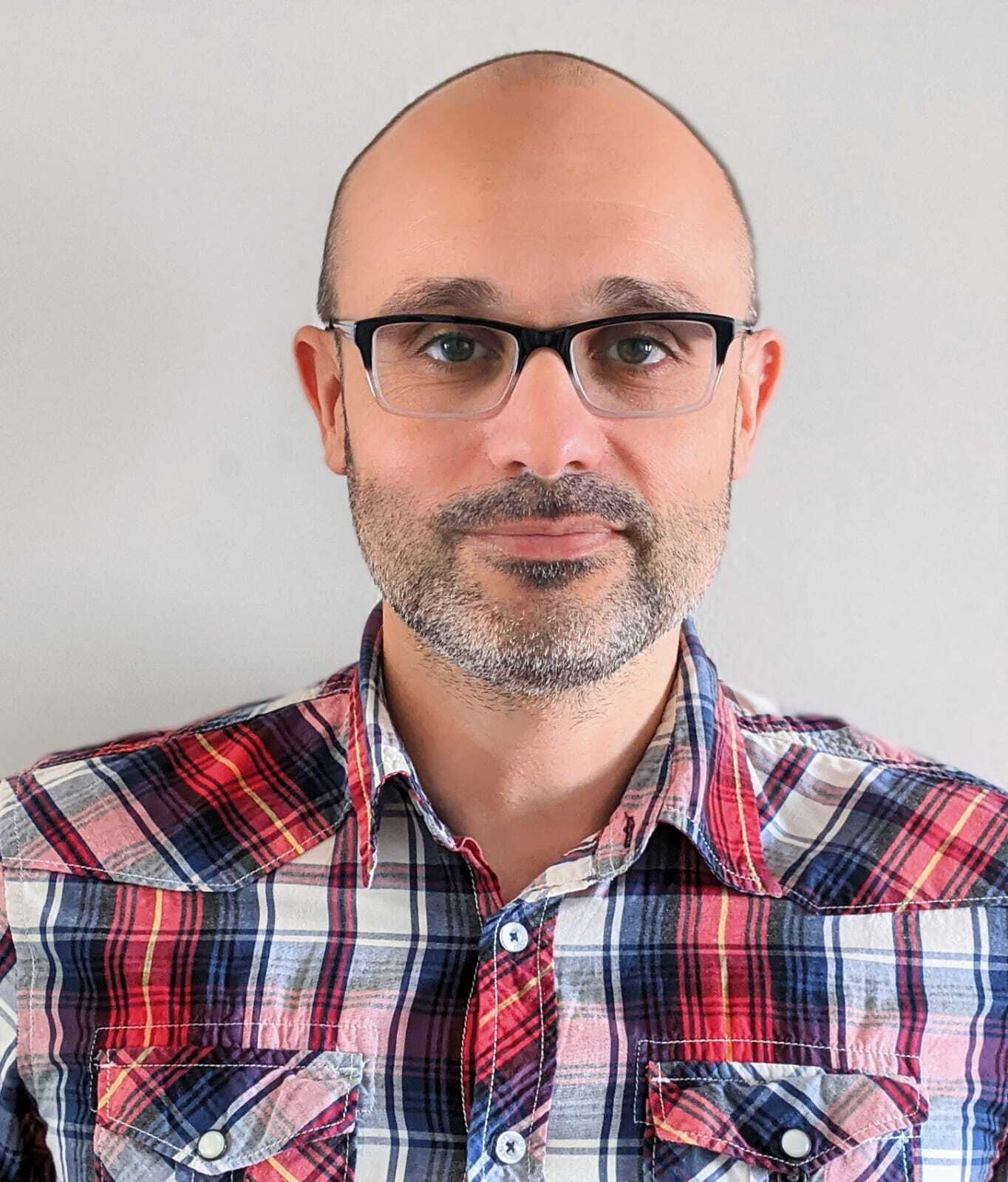}
\end{wrapfigure}\par

\noindent\textbf{Mirko Bronzi} is an Senior Applied Research Scientist at Mila-Québec AI Institute. He completed his Ph.D. in Rome (University of Roma Tre) working on techniques for extracting and integrating information from different web sources, in an automatic way. During his career he worked on NLU systems for virtual assistants for cars, on CLU (clinical language understanding), as well as in other domains of machine learning.
His research interests are in the NLP area and in the audio domain for machine learning. He is also interested in computer vision, and in the intersection of Engineering and machine learning.

\end{document}